\documentclass[sensors,article,accept,moreauthors,pdftex]{Definitions/mdpi} 

\usepackage{gensymb}
\usepackage{siunitx}
\usepackage{xcolor}
\usepackage[normalem]{ulem}
\firstpage{1} 
\pubvolume{1}
\issuenum{1}
\articlenumber{0}
\pubyear{2021}
\copyrightyear{2021}
\externaleditor{Communicated by: {Yang Yue}} %please add,  For journal Automation, please change Academic Editor to "Communicated by"
\datereceived{} 
\dateaccepted{} 
\datepublished{} 
\hreflink{https://doi.org/} % If needed use \linebreak
\usepackage[nolist]{acronym}
\begin{acronym}[MIMOOOO] % Give the longest label here so that the list is nicely aligned
\acro{MIMO}{Multiple-Input Multiple-Output}
\acro{LTE}{Long Term Evolution}
\acro{FFR}{Fractional Frequency Reuse}
\acro{ICIC}{Inter-Cell Interference Coordination}
\acro{SINR}{Signal-to-interference-plus-noise Ratio}
\acro{eNB}{e-NodeB}
\acrodefplural{eNB}{e-NodeBs}
\acro{eMBB}{enhanced Mobile Broadband}
\acro{ICI}{Inter-Cell Interference}
\acro{eICIC}{enhanced Inter-Cell Intereference Coordination}
\acro{QoS}{Quality of Service}
\acro{UE}{User Equipment}
\acrodefplural{UE}{User Equipments}
\acro{ABS}{Almost Blank Subframes}
\acro{CRE}{Cell Range Extension}
\acro{HetNet}{Heterogeneous Network}
\acro{OFDMA}{Orthogonal Frequency-Division Multiple Access}
\acro{BER}{Bit Error Rate}
\acro{mmWaves}{Millimiter Waves}
\acro{HCS}{Hierarchical cell structure}
\acro{CSG}{Closed Subscriber Group}
\acrodefplural{CSG}{Closed Subscriber Groups}
\acro{3GPP}{3rd Generation Partnership Project}
\acro{NoOp}{No Operation}
\acro{PLR}{Packet Loss Ratio}
\acro{RSRQ}{Reference Signal Received Quality}
\acro{RS}{Reference Signal}
\acro{RB}{Resource Block}
\acrodefplural{RB}{Resource Blocks}
\acro{OFDM}{Orthogonal Frequency-Division Multiplexing}
\acro{HeNB}{Home eNode B}
\acro{eICIC}{Enhanced ICIC}
\acro{feICIC}{Further Enhanced ICIC}
\acro{NR}{New Radio}
\acro{PRB}{Physical Resource Blocks}
\acrodefplural{PRB}{Physical Resource Blocks}
\acro{PRB}{Physical Resource Block}
\acro{EPC}{Evolved Packet Core}
\acro{E-UTRAN}{Evolved Universal Terrestrial Radio Access Network}
\acro{MME}{Mobility Management Entity}
\acro{S-GW}{Serving Gateway}
\acro{P-GW}{Packet Data Network Gateway}
\acro{RBG}{Resource Block Group}
\acrodefplural{RBG}{Resource Block Groups}
\acro{PDSCH}{Physical Downlink Shared Channel}
\acro{DL}{Downlink}
\acro{UL}{Uplink}
\acro{QL}{Q-Learning}
\acro{TTI}{Transmission Time Interval}
\acro{CDF}{Cumulative distribution function}
\acro{ANOVA}{Analysis of Variance}
\acro{ML}{Machine Learning}
\acro{RL}{Reinforcement Learning}
\acro{GppCom}{\textit{Grupo de Pesquisa em Prototipagem Rápida de Soluções para Comunicação}}
\acro{LTE-U}{LTE Unlicensed}
\acro{RRM}{Radio Resource Management}
\acro{MDP}{Markov Decision Process}
\acro{ARP}{Action-Replay Process}
\acro{AP}{Access Point}
\acrodefplural{AP}{Access Points}
\acro{PDU}{Protocol Data Unit}
\acrodefplural{PDUs}{Protocol Data Units}
\acro{RLC}{Radio Link Control}
\acro{HFR}{Hard Frequency Reuse}
\acro{Strict FR}{Strict Frequency Reuse}
\acro{SFR}{Soft Frequency Reuse}
\acro{SFFR}{Soft Fractional Frequency Reuse}
\acro{NoOp}{Full Frequency Reuse}
\acro{ns-3}{network simulator 3}
\acro{SS}{Sum of Squares}
\acrodefplural{SS}{Sum of Squares}
\acro{CRB}{Cell Range Bias}
\end{acronym}
\usepackage[english, ruled, linesnumbered]{algorithm2e}
\usepackage{booktabs,subcaption,dcolumn}
\newcolumntype{d}[1]{D..{#1}}
\Title{Solution for Interference in Hotspot Scenarios Applying Q-Learning on FFR-Based ICIC Techniques}

\TitleCitation{Solution for Interference in Hotspot Scenarios Applying Q-Learning on FFR-based ICIC Techniques}

\Author{\highlighting{Iago Diógenes do Rego} %Please carefully check the accuracy of names and affiliations. Changes will not be possible after proofreading.
 *\orcidA{} and Vicente A. de Sousa, Jr. \orcidB{}} %Please carefully check the accuracy of names and affiliations. 

\AuthorNames{Iago D. do Rego and Vicente A. de Sousa}

\AuthorCitation{Diógenes do Rego, I.; de Sousa, V.A.}

\address[1]{%
Department of Communications Engineering, Federal University of Rio Grande do Norte, \linebreak~Natal 59078-970, Brazil;  vicente.sousa@ufrn.br 
}

\corres{\hangafter=1 \hangindent=1.05em \hspace{-0.82em}Correspondence: iago.diogenes.072@ufrn.edu.br}
\abstract{This work explores interference coordination techniques (inter-cell interference coordination, ICIC) based on fractional frequency reuse (FFR) as a solution for a multi-cellular scenario with user concentration varying over time. Initially, we present the problem of high user concentration along with their consequences. Next, the use of multiple-input multiple-output (MIMO) and small cells are discussed as classic solutions to the problem, leading to the introduction of fractional frequency reuse and existing ICIC techniques that use FFR. An exploratory analysis is presented in order to demonstrate the effectiveness of ICIC techniques in reducing co-channel interference, as well as to compare different techniques. A statistical study was conducted using one of the techniques from the first analysis in order to identify which of its parameters are relevant to the system performance. Additionally, another study is presented to highlight the impact of high user concentration in the proposed scenario. Because of the dynamic aspect of the system, this work proposes a solution based on machine learning. It consists of changing the ICIC parameters automatically to maintain the best possible signal-to-interference-plus-noise ratio (SINR) in a scenario with hotspots appearing over time. All investigations are based on ns-3 simulator prototyping. The results show that the proposed Q-Learning algorithm increases the average SINR from all users and hotspot users when compared with a scenario without Q-Learning. {The SINR from hotspot users is increased by 11.2\% in the worst case scenario and by 180\% in the best case.}}

\begin{document}

\section{Introduction}

According to a Cisco forecast \cite{Cisco2011}, by 2022, traffic from wireless and mobile devices will account for 71\% of global IP traffic. Between 2017 and 2022, all data traffic in mobile networks will be seven times bigger, as a result of the intense sharing and consumption of data, especially video. Recent studies \cite{gsma2019} predict that, over the next 7 years, 1.4 billion people will start using mobile internet for the first time. Until 2025, over 60\% of the world's population will be using mobile internet. Additionally, there is an ongoing change of consuming habits. As more people consume video-related content, longer and more frequently, especially via mobile devices, the use of mobile data per user will be five times bigger by 2024.

Nonetheless, smartphones are not the only devices responsible for the increase on traffic demand. According to the GSMA Association \cite{gsma2019}, between 2018 and 2025, the amount of IoT devices in mobile networks will triple, reaching 25 billion connections. Furthermore, 5G will also increase the number of connected devices and use cases offered by the network. This constant growth of traffic demand calls for higher network capacity. In fact, the Shannon--Hartley theorem \cite{Shannon1998} demonstrates that increasing the available bandwidth is the most effective way to increase a channel's capacity. However, the spectrum is a limited resource that has to be used efficiently.

In order to increase network capacity, the evolution of wireless communication technologies has introduced approaches, such as the use of multiple antennas, power control, management of co-channel interference by reusing the spectrum, and the implementation of more efficient modulation schemes. For example, \ac{LTE} originally proposes a reuse factor of 1 in order to increase spectrum efficiency. A reuse factor of 1 means that all network cells operate in the same frequency band. A reuse factor of 3, on the other hand, decreases the band available to each cell by 3, decreasing the channel's capacity. However, a reuse factor of 1 may lead to low \ac{SINR}, especially to users located further away from the \ac{eNB} (radio base station on \ac{LTE}), due to interference from neighboring cells.

These techniques aim at increasing channel capacity, but they also introduce new challenges, mainly related to the compromise between coverage and quality of service. For example, \ac{eMBB} has been a widely discussed topic {in} the literature. It represents the evolution of mobile communications and it has been the focus in the first commercial deployments of 5G. One of its goals is to provide high data rates and wide coverage. However, as 5G operates in high frequencies, the coverage tends to decrease, as the signal attenuation is more severe. Consequently, the network will likely have to deploy more base stations, increasing the need to mitigate interference from adjacent cells.

{In addition, some aspects that have strong impacts on mobile network performances, such as the traffic demand or the amount and distribution of users, are often unpredictable or variant through time.} Big cities usually host events that can last from a few hours to a couple of days and these can be either recurrent or a one time occasion. Events and locations, such as school fairs, football games, music festivals, school parades, food parks, and malls can suddenly increase the number and density of users at that location.

The planning and deployment of a mobile network is traditionally associated with modeling the channel based on the specific demand for a location. This usually does not take into account some parameters that can vary in an unexpected way, as the ones mentioned above. In non-dynamic systems, the occasional appearance of areas with high user concentration can severely degrade the channel quality and the system's ability to serve its users. 

In this work, the term hotspot is used to describe this region with a high density of users and it does not mean a new \ac{AP}. Besides the increase on traffic demand, due to a high number of users, there is also a higher probability of increased interference, especially on systems with a reuse factor of 1. In {such} scenarios, the users in the borders of the coverage area are the most affected by interference.

The paper is organized as follows. Section \ref{sec:interferenceHotspot} presents the problem of high user concentration and its impact on system performance, along with some classical solutions for this problem. Sequentially, Section \ref{sec:ffr} introduces \acf{FFR}, an important component of the proposed solution, and Section \ref{sec:icic_3gpp} shows how \ac{3GPP} has developed \ac{ICIC} techniques over the years. Section \ref{sec:related-works} presents relevant related works and Section \ref{sec:systemModel} indicates how the system is modeled for simulations. Finally, Sections \ref{sec:performanceComparison}--\ref{sec:hotspot} show results from {preliminary} analysis and Section \ref{sec:solution} presents our proposed solution, as well as the proof-of-concept results and discussion.

In this paper, our key contributions are:
\begin{itemize}
    \item Performance comparison of classic FFR-based ICIC algorithms that take into account the number of users and their distributions along the cell (edge and center);
    \item Statistical analysis of the strict frequency reuse scheme that identifies which parameters do not need to be dynamically controlled;
    \item Analysis of {the strict frequency reuse algorithm's performance} in a mixed scenario with hotspot and homogeneous user distribution; 
    \item Q-Learning algorithm that continually operates in the network to dynamically mitigate the performance loss (SINR) that results from the appearance of hotspots (densely populated areas).
\end{itemize}

%%%%%%%%%%%%%%%%%%%%%%%%%%%%%%%%%%%%%%%%%%
\section{Interference in Hotspot Scenarios}\label{sec:interferenceHotspot}

Mobile networks face many challenges that can limit channel capacity. For example, the lack of available bandwidth is a common issue that can be addressed by increasing the spectrum reuse, as a way to increase the capacity limit. However, that may lead to a higher co-channel interference, which happens when more then one user is using the same radio resource at the same time.

The solution presented in this paper can be applied on an \ac{OFDMA}-based system (e.g., 4G LTE or 5G NR) in a \ac{DL} direction, whose users may experience interference from multiple neighboring cells reusing the same frequency at the same time. This interference may be a severe limiting factor in terms of capacity, especially with a reuse factor of 1.

In such scenarios, users from different cells can interfere with each {other, and} users located in the cell edge are the most affected by co-channel interference~\cite{Zheng2015}. These users are far from the transmitting cell and closer to adjacent cells, increasing the probability of receiving any interfering signals with higher power levels and signals from its serving cell with lower power levels. Hence, user distribution is another aspect that can have {a} negative impact on system performance.

Therefore, the existence of regions with high user concentration (hotspots), {especially} if close to the cell edge, can lead to {a} lack of available resources and {a} higher probability of users being allocated to the same frequency. Regardless {of} its location, the appearance of hotspots may increase interference, leading to lower \ac{SINR}. This results in higher \ac{BER} and, consequently, lower traffic capacity.

All of these challenges are well known and can degrade performance; however, they are partially predictable. On the other hand, some modern, urban hotspot scenarios bear a more dynamic aspect. For example, the {number} of people flowing through a big city {does} not merely repeat weekly, since extraordinary situations that are hard to predict often happen in large urban areas.

For instance, a certain neighborhood that consistently had a uniform distribution of users may change significantly because a new food park has opened, a soccer league changed its venue, a protest is taking place or an accident stopped the traffic. In such situations, mobile networks that were planned and deployed statically will not be able to serve {its users well}. Besides, the traffic demand per user is also likely to increase as these situations or events induce a different behavior in users. It leads them to consume more bandwidth by sharing photos and videos, {aggravating} the high traffic demand.

\subsection{Classic Solutions for Interference in Hotspot Scenarios}\label{sec:classicSolutions}

\subsubsection{MIMO Systems}

\ac{MIMO} is a technology that uses multiple antennas for transmitting and receiving signals {to} mitigate negative aspects of the channel and/or multiplex data transmission. The first goal, for example, can be attained by increasing the diversity of transmitted signals \cite{livro-vicente}. The unpredictability of the wireless channel is used as a tool to improve system performance \cite{Diggavi2004} by using multiple antennas to exploit the advantages of spatial diversity \cite{Adhikary2013}. Furthermore, MIMO can also explore spatial multiplexing, {using multiple} antennas to transmit data through various channels.

Given a minimum spacing between antennas, it is possible to obtain independent or weakly correlated channels on each Tx/Rx (transmission and reception) pair. This concurrent transmission in the same frequency band can increase data rates or improve reliability without compromising spectral efficiency.

A scenario with hotspots is prone to have some channels suffering with less interference {than} others. Thus, if an appropriate combination method is applied for the different transmissions, it is possible to improve performance. For example, choosing the path with better \ac{SINR} \cite{livro-vicente} is a simple solution that can lead to better results. Additionally, multiple antennas can also control the signal amplitude and phase to direct beams towards certain users or regions and avoid users suffering from interference \cite{Adhikary2015}.

Recent wireless communication systems point \ac{MIMO} as an efficient approach to increase data rates, {especially} when suffering from {small-scale} multi-path, such as systems with hotspots. For instance, the use of multiple antennas has been widely discussed in the literature as a core feature of 5G systems \cite{Muirhead2016}, through massive MIMO. The use of \ac{mmWaves} enables antenna arrays with a high number of elements \cite{Yaacoub2016}. However, MIMO is a solution implemented in the physical layer, which requires investments {in} hardware and it often demands channel estimation.

Nevertheless, this work does not intend to {indicate} MIMO as a technique to be replaced. {Still, it} introduces \ac{FFR} as an appropriate technology to address the specific problems noted {in} the last section. In mobile networks, \ac{MIMO} often co-exists with \ac{FFR} techniques.

\subsubsection{Small Cells}

\acf{HetNet} is an extended concept of \ac{HCS} that has been discussed even before the standardization of \ac{LTE} \cite{lte-hetnet}. It is usually associated with urban scenarios, where the traffic demand and the number of users {are} constantly increasing.

In \acp{HetNet}, different categories of cells coexist with each other, and they usually differ in coverage and capacity \cite{Damnjanovic2011}. {Still, }they do not necessarily need to share the same radio access technology. In such deployments, small cells are an alternative to offload the macrocells traffic, especially on hotspots and cell-edge \cite{dynamic_hetnet}, due to severe interference.

Dense urban areas that already have limited capacity may require big investments to deploy another conventional access point. In such cases, the use of small cells is attractive, since they can improve coverage and transmit with lower power, given the smaller distance between user and transmitter. These cells usually operate in the same frequency band as the macrocells, transmitting with lower power to reduce interference.

Recent studies involving \acp{HetNet} usually focus on two categories of small cells: femtocells and picocells \cite{survey-femtocell,Al-Turjman2019,icic-lte-a-survey}. The former is typically deployed by the user, it is not planned, and it may be private, being called a \ac{CSG}. The latter has the functionality of a regular \ac{eNB}, but with smaller coverage and lower transmission power. Picocells are usually deployed to cover areas with {a} high density of users and they can also be deployed indoors, if necessary.

The use of small cells is also indicated as a key technology for 5G. The densification of the network has been indicated in the literature as an efficient solution to help satisfy the requirements for 5G \cite{5g-small-cells}.

However, the use of small cells also imposes some challenges. Its deployment demands investment in planning and hardware, and it may increase the amount of unnecessary hand-offs~\cite{Chang2017}. The large power difference between cell tiers can leave small cell users {at a} disadvantage. On the other hand, \acp{CSG} can create holes in coverage for the macrocell users, as they do not have access to the \ac{CSG}.

\section{Fractional Frequency Reuse}\label{sec:ffr}

A mobile network usually offers coverage to its subscribers with cells that have a limited frequency band for transmitting and receiving signals. In order to use the spectrum efficiently, a reuse factor of 1 is commonly applied, which means that all of the cells use the entire available bandwidth.

\begin{figure}[H]

%\vspace{0.5cm}
\includegraphics[width=0.6\linewidth]{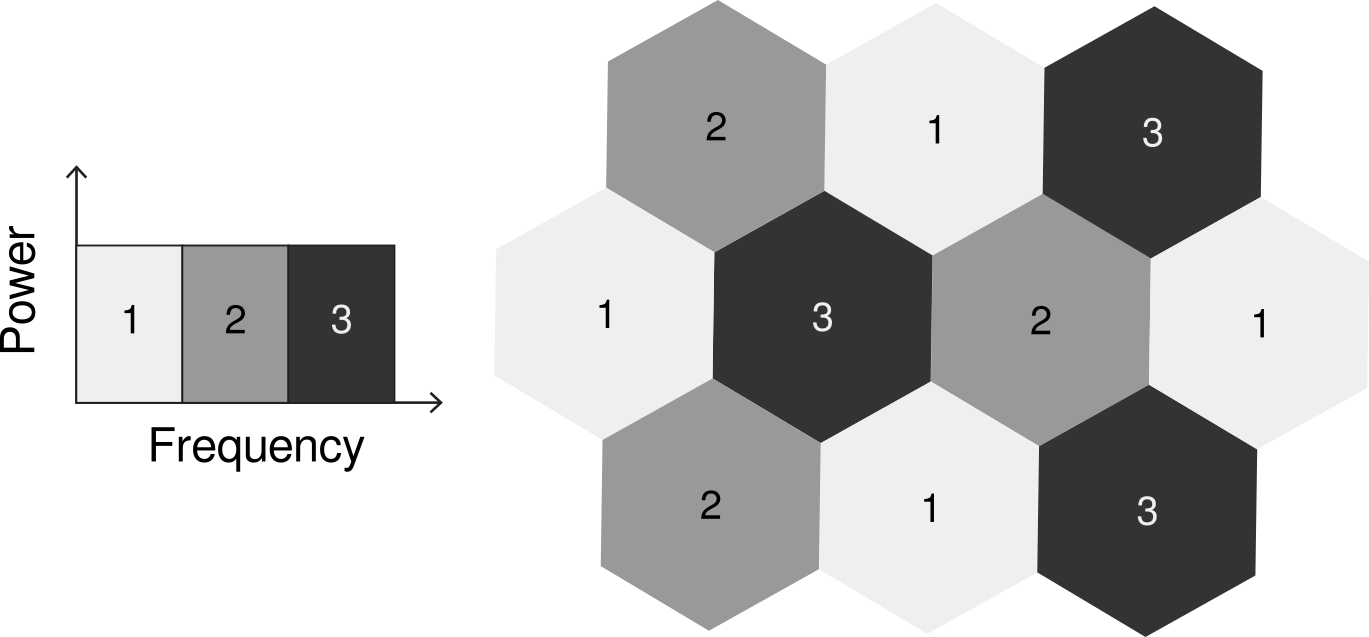}
\caption{\label{fig:reuse3} {Scenario with reuse factor of 3. In this topology, no adjacent cells share bandwidth, i.e., their frequency bands are disjoint.}}
\end{figure}

In order to mitigate inter-cell interference, some networks increase the reuse factor. For example, a reuse factor of 3 creates a topology where adjacent cells do not share bandwidth, as illustrated in Figure \ref{fig:reuse3}. However, there is a significant loss in spectral efficiency, given that only 1/3 of the original bandwidth is available to each cell.

Fractional frequency reuse splits the cell into distinct regions with different reuse factors and transmission power. The goal is to improve the \ac{SINR}, increasing the reuse factor without compromising the spectral efficiency. Users in the cell edge are usually the target of FFR-based techniques, since they are the most affected by co-channel interference.

The {following} section presents some FFR-based techniques, known in the \ac{3GPP} as \ac{ICIC} techniques. These algorithms were selected because of the following reasons. They are standardized for \ac{LTE} by \ac{3GPP} (but not limited to it, since they can be applied to other OFDMA-based systems, such as 5G NR), they are already implemented in the simulator used in this paper \cite{baldo}, and they represent a good range of \ac{ICIC} strategies based on FFR that can be found in deployed networks \cite{khalifa}. The algorithms are fully described in \cite{khalifa,kimura,eFFR}.

\subsection{ICIC Techniques}\label{sec:icic-techniques}

In order to compare the techniques, two algorithms without fractional reuse of the frequency are also considered: \ac{NoOp} and \ac{HFR}. The former has a reuse factor of 1, the latter has a reuse factor of 3, as illustrated {in} Figure \ref{fig:reuse3}, and both transmit with the same power for the whole bandwidth.

In scenarios with few users, the \acf{NoOp} may present high data rates given the larger available bandwidth for each cell. However, higher interference is expected, and it may result in low \ac{SINR} and high \ac{PLR}, {especially} for edge users. The \ac{HFR} algorithm is more efficient {in} reducing interference, as all adjacent cells have disjoint bandwidths. However, each cell has only 1/3 of the available bandwidth, which can severely reduce throughput, depending on the offered load \cite{kimura}.

\subsubsection{Strict Frequency Reuse}

\begin{figure}[H]

\includegraphics[width=0.6\linewidth]{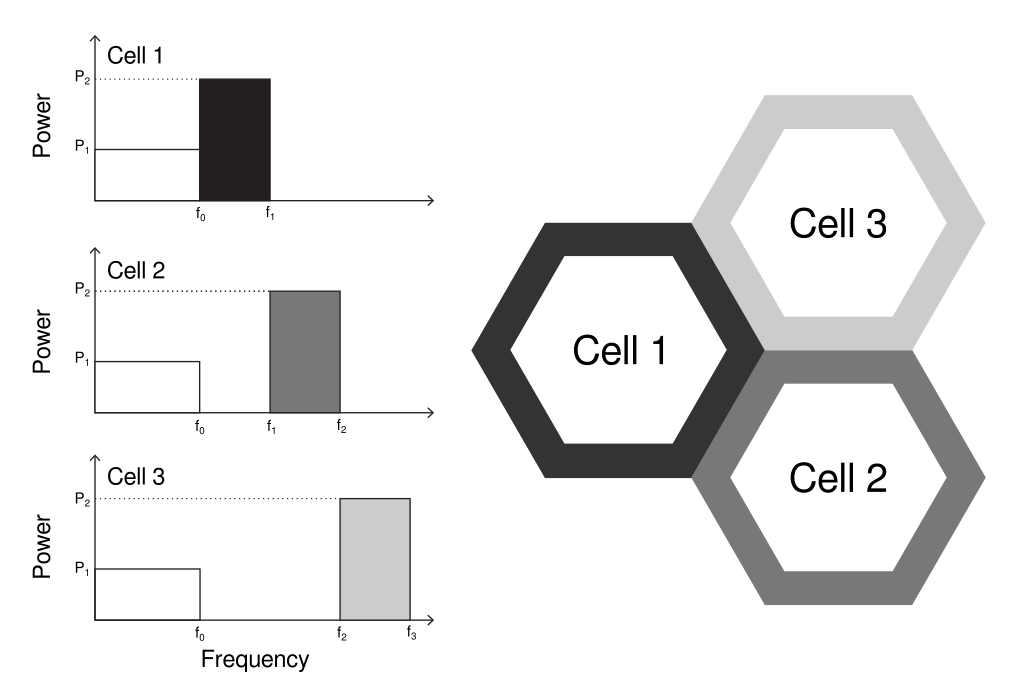}
\caption{\label{fig:strict} {Bandwidth distribution and power allocation for the Strict frequency reuse algorithm in a cluster of three cells.}}
\end{figure}

The \ac{Strict FR} splits the bandwidth into two sub-bands, as illustrated {in} Figure \ref{fig:strict}. The cell-center \acp{UE} are allocated to a common sub-band shared by all cells (reuse factor of 1) and the cell-edge \acp{UE} are allocated to a private sub-band (reuse factor of 3). Hence, cell-center users share bandwidth with neighboring cells, thus increasing spectral efficiency, but cell-edge users do not, thus reducing interference. Additionally, a higher power level is used for the private sub-band.

In order to determine whether a \ac{UE} is allocated to the common or private sub-band, the \ac{ICIC} algorithm uses a metric defined in \ac{3GPP}, the \ac{RSRQ}. It indicates the quality of the signal and it takes into account various metrics, such as noise, power from interfering signals, and the number of allocated \acp{RB}. If the \ac{RSRQ} reported by a user is higher {than} a threshold, the \ac{UE} is allocated to the common sub-band (cell-center). Otherwise, it is allocated to the private sub-band (cell-edge). The \ac{RSRQ} threshold is {a} user-defined parameter.

\subsubsection{Soft Frequency Reuse}

\begin{figure}[H]
\includegraphics[width=0.6\linewidth]{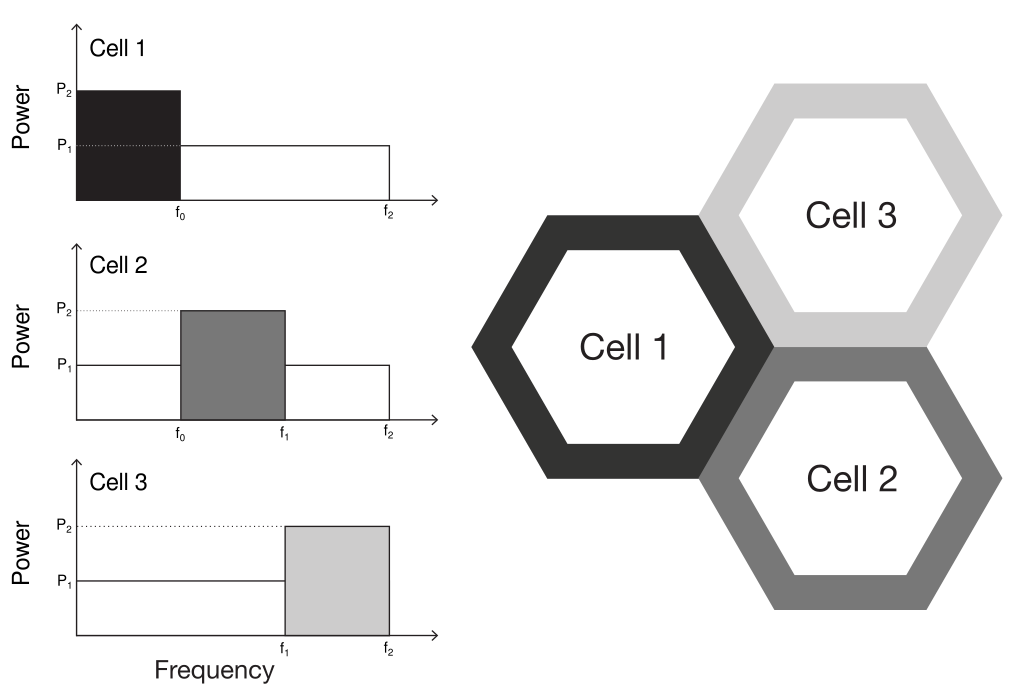}
\caption{\label{fig:soft} {Bandwidth distribution and power allocation for the soft frequency reuse algorithm in a cluster of three cells.}}

\end{figure}

The \ac{SFR} also splits the bandwidth into two sub-bands. However, the sub-band allocated to cell-edge \acp{UE} is not private, as illustrated {in} Figure \ref{fig:soft}. However, cell-edge users only share bandwidth with cell-center users from adjacent cells. Consequently, the cell-edge sub-bands are disjoint. Thus, cell-edge \acp{UE} do not use the same frequency band as neighboring cells, and each occupies 1/3 of the available spectrum.

The \ac{SFR} can lead to better spectral efficiency when compared to the \ac{Strict FR}, given that all cells can use the entire bandwidth. However, the \ac{SFR} also increases the interference suffered by all users.

A higher power level is used for the edge sub-band, and a \ac{UE} is considered in the cell-center if the reported \ac{RSRQ} is greater than the threshold. Otherwise, it is allocated in the cell-edge sub-band.

\subsubsection{Soft Fractional Frequency Reuse}

\begin{figure}[H]
\includegraphics[width=0.6\linewidth]{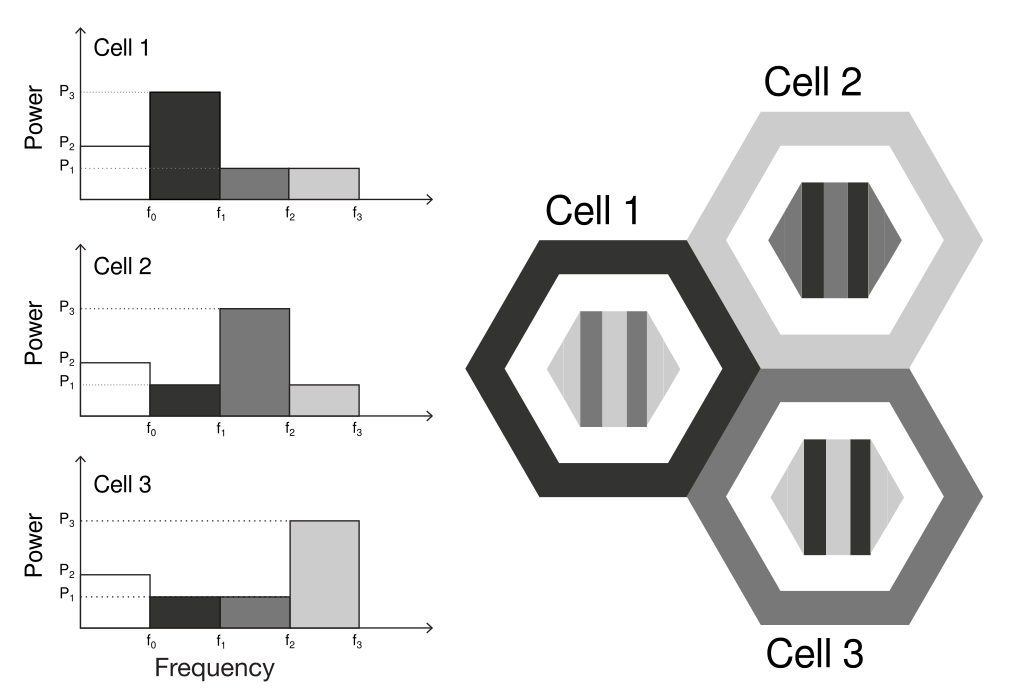}
\caption{\label{fig:sffr} {Bandwidth distribution and power allocation for the soft fractional frequency reuse algorithm in a cluster of three cells.}}

\end{figure}

The last technique is the \ac{SFFR}. It divides the bandwidth {into} three distinct sub-bands: center, middle, and edge, as illustrated on \mbox{Figure \ref{fig:sffr}}. The middle sub-band has a reuse factor of 1 and the edge sub-band has a reuse factor of 3. The center region reuses the frequency band from the edge of the adjacent cells.

Each sub-band is served with a different power level, increasing from center to edge, and user allocation on sub-bands is also done based on the reported \ac{RSRQ}.

\section{ICIC in 3GPP standards}\label{sec:icic_3gpp}

\ac{LTE} is a wireless communication standard introduced by \ac{3GPP} on Release 8. It has an all-IP network architecture and high flexibility on spectrum allocation, since \ac{OFDMA} is the multiple access scheme on \ac{DL}. Consequently, the \ac{eNB} can allocate a \ac{UE} to any sub-carrier in the frequency domain. This allocation is based on \acfp{RB} and each \ac{RB} has 12 sub-carriers and 180~kHz of minimal bandwidth \cite{livro-umts-lte-a}. This flexibility is important to enable fractional reuse of the spectrum.

The X2 interface, introduced on \ac{LTE}, is also important for \ac{ICIC}. It enables signaling between \acp{eNB}, allowing for the \ac{ICIC} algorithms to manage the \ac{RB} and power allocation on neighboring cells \cite{livro-lte-a-3gpp}.

Moreover, \acs{3GPP}'s Release 8 also introduces implicit support to interference coordination \cite{ts36213}, given that \ac{LTE} has good control over time, frequency, and power resources. Release 9 presents studies that attempt to mitigate interference between macrocell and lower power nodes and, further on, Release 10 extends \ac{ICIC} to the time domain as it introduces \ac{eICIC}. Additionally, Release 11 presents the \ac{feICIC} in order to mitigate interference due to control signals.

Given this history, \ac{eICIC} and \ac{feICIC} are often considered as the evolution of \ac{ICIC}. However, FFR-based techniques are still relevant based on recent studies, given that \acf{NR}, the new \ac{3GPP} standard for 5G systems, demands a flexible and efficient use of the spectrum \cite{ts-5g-nr}. For example, the authors of \cite{Soret2018} propose a FFR-based \ac{ICIC} scheme for \ac{NR}. The algorithm dynamically allocates users suffering from interference to a set of \acp{PRB} that are not accessible to interfering cells.

\section{Related Works}\label{sec:related-works}

Metropolitan areas around the world have been growing significantly, alongside the number of users on wireless networks. This growth has encouraged the literature to widely discuss scenarios with {a} high density of users. Hence, the densification of the network is indicated as an efficient approach to better serve urban areas, including 5G networks \cite{Al-Turjman2019}. Consequently, various works introduce ways of using small cells to boost the performance of scenarios with hotspots.

The authors of \cite{Singh2018} propose a scenario where macrocells are populated with picocells centered on hotspots. The paper shows that the small coverage offered by the picocells causes an imbalance {in} user allocation, overloading the macrocells. Therefore, a dense ring of macrocell users is formed around the picocells. These users experience high interference coming from the picocells, and they are considered as victim users. They propose two dynamic techniques of \ac{eICIC} using two different metrics to improve the \ac{QoS} of the victim users while muting picocell users.

A realistic dense scenario was evaluated in \cite{Lopez-Perez2015}. It simulated a wide urban area with 300 active users per $km^2$. Half of the users were uniformly distributed withon the scenario, and the other half were uniformly distributed inside circular hotspots with a 40~m radius. For this scenario, the simulation results show that it is possible to achieve an average of 1 Gbps per user. However, they also noted that, given the year of publishing, the solution was not applicable due to its high cost and low energy efficiency.

The authors of \cite{Shirakabe2011} evaluated a scenario that deployed macrocells with three sectors each and four picocells per sector. Moreover, a fraction of the users {were} allocated in hotspots centered in the picocells. They proposed the use of time-domain \ac{ICIC}, using \ac{ABS} and \ac{CRE} to improve performance and alleviate the macrocell load. The results show that the imbalance between cells is effectively countered with an offset value of 10 dB for the \ac{CRE}. Besides, for each \ac{CRE} offset, there was an optimum value for the ratio of protected subframes. Thus, if both parameters are properly configured {within a certain range}, the performance is almost the same.

Another technique used to improve {the} performance of hotspot scenarios is the use of multiple antennas. The authors of \cite{Adhikary2015} proposed a \ac{HetNet} scenario with small cells operating in the same frequency as macrocells. All users were allocated in hotspots and the small cells were located at the center of some of these hotspots, representing an intentional deployment of small cells. The solution focuses on mitigating the interference coming from the small cells through a beamforming scheme proposed in \cite{Adhikary2013}. This scheme concentrates the transmission energy to the hotspots while creating transmission opportunities for users {in} other directions.

A scheduling algorithm that combines frequency allocation and beamforming (beams width and direction) is proposed in \cite{Shen2017}. A homogeneous and a hotspot scenario were considered. The paper focused on maximizing throughput according to \ac{QoS} requirements, and it compared the results to other approaches. The results show that the hotspot scenario is more challenging for all the algorithms. However, the proposed solution has better performance in terms of complexity and throughput.

The authors of \cite{Ruegg2016} proposed a solution that served \acp{UE} in a hotspot scenario using virtual \ac{MIMO}, i.e., user cooperation to enable spatial multiplexing. Compared to the traditional use of \ac{MIMO}, this approach can avoid the costs of new antennas or access points dedicated to the hotspots. Furthermore, the signal processing was done in the mobile station, avoiding any new processing units. The proposed protocol presents better performance {than} traditional offloading techniques, but there is no discussion about privacy, the impact of the added signaling, or energy consumption on the user's side.

The literature has widely discussed solutions to the problems introduced by small cells, but, in general, only a few objectively show the negative effects related to the appearance of hotspots. In general, the deployment of small cells and the use of \ac{MIMO} techniques demand investments {in site-planning} and infrastructure, and the latter usually requires channel estimation.

\ac{FFR} is still relevant as an efficient approach to mitigate inter-cell interference in mobile networks \cite{Khan2019a,Li2020,Zheng2019}. Besides, recent studies still point out that \ac{FFR} can be combined with different techniques to reduce interference, such as beamforming \cite{Li2020}. 

Hotspot scenarios are consistently studied for {their} importance regarding current and future mobile networks \cite{Mitsolidou2019, Ma2020}. {Still,} no work has been found in the use of \ac{FFR} as the {primary} solution to improve performance on hotspot scenarios without using small cells. \ac{FFR} can be a good alternative for its simplicity and efficiency, even in heterogeneous scenarios with high user density \cite{Abdullahi2017}. This is one of the arguments that motivates the scientific hypothesis of this work.

\section{System Model}\label{sec:systemModel}

\subsection{Simulation Software}

The \acf{ns-3} is the simulation tool used in this work~\cite{ns-3}. It is open-source and, therefore, publicly available for development, education, and research activities.
It is also modular, comprised of several models built primarily in C++ with some APIs available in Python. Its development is oriented by technical specifications from standard organizations, such as \ac{3GPP} and IEEE. Besides, it is well documented, with an active and collaborative community. For these reasons, it has been widely adopted for research and is the {primary} simulation tool for this paper.

\subsection{\acs{LTE} Module and \acs{ICIC} on ns-3}\label{sec:lteModule}

\begin{figure}[H]
\includegraphics[width=0.7\linewidth]{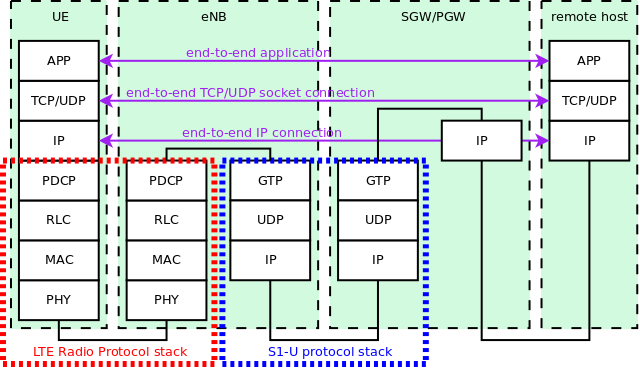}
\caption{\label{fig:lte-epc} {LTE-EPC data plane protocol stack (taken from the ns-3 documentation \cite{lte-model}).}}
%\vspace{0.45cm}

\end{figure}

We provide our proof-of-concept results using {an} LTE system model. The \ac{LTE} module on \ac{ns-3} has two main components: the \ac{LTE} model and the \ac{EPC} model. The former includes the \acs{E-UTRAN} protocol stack, i.e., the RRC, PDCP, RLC, MAC, and PHY layers. These entities reside within the \ac{UE} and the \ac{eNB} nodes. The \ac{EPC} model includes the core network functionalities, allowing end-to-end IP connectivity. All protocols and entities reside within the \acs{MME}, \acs{S-GW}, and \acs{P-GW} nodes and partially within the \ac{eNB} nodes. Figure~\ref{fig:lte-epc} shows the LTE-EPC protocol stack for the data plane on ns-3~\cite{lte-model}. The only relevant simplification is the combination of the \acs{S-GW} and \acs{P-GW} functionalities into \mbox{one node.}

There are currently seven \ac{ICIC} algorithms implemented in the \ac{LTE} module, described in \cite{A.S.HamzaS.S.KhalifaH.S.HamzaandK.Elsayed2013}. The \ac{FFR} algorithms act on scheduling, commanded by the MAC layer \cite{lte-model}. The algorithm is consulted and, depending on its rules for bandwidth allocation, it may allow (or not) the scheduling of a \ac{UE} to a certain \ac{RBG}. These algorithms have three main parameters that can be configured, as described below.

The power level of a sub-band is defined through a power offset between the \ac{RS} and the \ac{PDSCH}. Each sub-band has a variable that defines this offset in decibels. For example, the \ac{Strict FR} has \textit{CenterPowerOffset} and \textit{EdgePowerOffset}.

Resource allocation in \ac{LTE} is done through \acsp{RBG}, but its size in number of \acp{RB} depends on the system bandwidth \cite{ts36213}. For example, for a system bandwidth of 100 \acp{RB}, each \ac{RBG} has four \acp{RB} \cite{ts36213}. Hence, each sub-band has a variable that defines the number of available \acp{RBG} for \ac{DL} and \ac{UL}, separately.

Section \ref{sec:icic-techniques} introduced the \ac{RSRQ} threshold, which determines user allocation on sub-bands. On \ac{LTE}, \ac{RSRQ} is measured in decibels and mapped {into} integer values before being reported \cite{ts36133}. On Release 8, these values range from 0 to 34. Hence, the variable \textit{RsrqThreshold} only assumes values within this range.

\subsection{Q-Learning on ns-3}

In order to execute the proposed solution, a new class was added to ns-3. It implements the \acf{QL} algorithm, which will be described in Section \ref{sec:solution}. Its first version was introduced by the authors of \cite{gppcom-dm-csat}.

The simulation script includes both models described in Section \ref{sec:lteModule}. The \ac{EPC} model enables the installation of applications in the \acp{UE}, which allows better control over the offered load and the appearance of hotspots during the simulation, {considering} the applications can be turned on and off at any given time.

All calculations related to the \ac{QL} algorithm are made during \textit{runtime}, while the simulation is executed. The algorithm also operates in the system during \textit{runtime}, which guarantees that the network is able to adapt dynamically.

\section{{Preliminary Analysis A: Performance Comparison of ICIC Algorithms}} \label{sec:performanceComparison}

{The proposed solution of this paper is fully described in Section 10. However, before presenting the final results, we present three preliminary analyses that were important to guide some decisions regarding the Q-Learning algorithm. The first analysis is presented in this section, and} it is a performance comparison between the \ac{ICIC} techniques presented in Section~\ref{sec:icic-techniques}. This study evaluates the impact of \ac{FFR} on cell-edge \acp{UE} in a simple scenario. The algorithms were executed under the same conditions and {their} parameters assume the default values defined either on ns-3 or in the literature.

\subsection{Evaluation Scenario}\label{sec:scenarioPerformanceComparison}

\begin{figure}[H]
\includegraphics[width=0.55\linewidth]{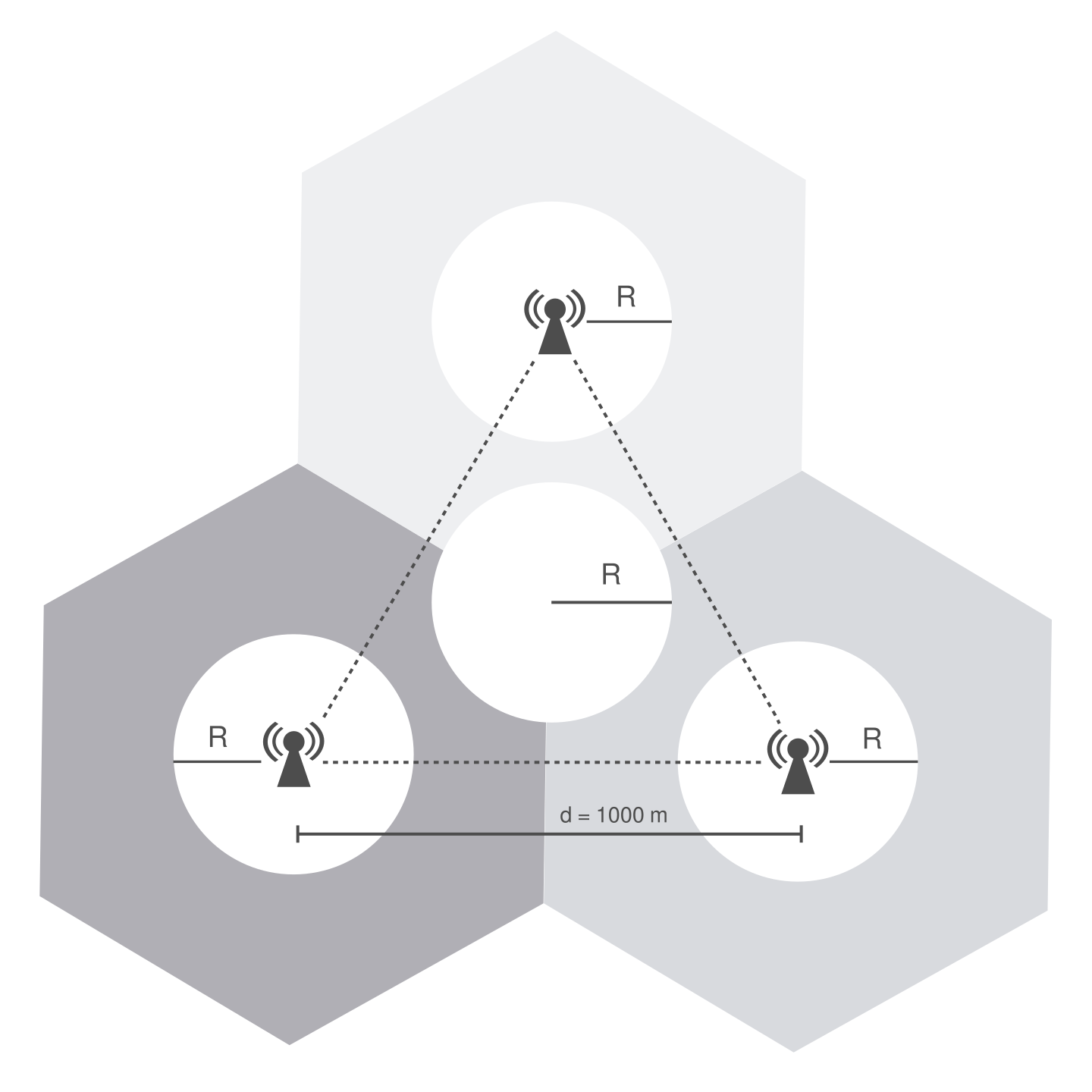}
\caption{\label{fig:scenarioPerformanceComparison} {Scenario for preliminary analysis a: performance comparison of ICIC algorithms. The Radius R is set to 100 m in Scenario 1 and 500 m in Scenario 2, as described in Section \ref{sec:scenarioPerformanceComparison}.}}
\vspace{0.25cm}

\end{figure}

The scenario consists {of} 3 \acp{eNB} positioned in the vertices of an equilateral triangle with side equal to 1000 m, as illustrated {in} Figure \ref{fig:scenarioPerformanceComparison}, and each \ac{eNB} is the center of a cell. This scenario was introduced by the authors of \cite{baldo} to evaluate the \ac{ICIC} techniques implemented on ns-3.

Users are distributed uniformly inside circles controlled by a radius that can be configured as needed. There is a circle centered on each \ac{eNB} and an additional circle in the triangle's centroid, which is the farthest location from each \ac{eNB}.

User location within the circles is randomly selected from {a} uniform distribution, and all circles have the same radius. Given that a smaller radius leads to a more densely populated region (hotspot), it is possible to control user concentration. The {number} of users vary from 10 to 60 \acp{UE} on each circle and different circles always have the same amount of users. Namely, if there are 30 users in the {cell edge} area, there are also 30 users on each cell-center, i.e., 120 in total. Therefore, the {number} of users range from 40 to 240.

Scheduling is made using the proportional fair algorithm and the link adaptation is based on \ac{SINR} for a \ac{BER} of $5\cdot10^{-5}$. The system bandwidth is 5~MHz, resulting in 25 \acp{RB} for each \ac{TTI}. {\acp{TTI} are the parameters related to the medium access in the radio link layer. At each \ac{TTI}, the scheduling algorithm allocates \acp{RB} for all connected users. It is a parameter strongly related to access latency on \ac{LTE} systems.}. Table \ref{tab:icicparamPerformaceComparison} presents the configuration for each \ac{ICIC} algorithm, suggested in \cite{baldo}. Other simulation parameters are presented in Table \ref{tab:paramPerformaceComparison}.

The simulation was conducted in two scenarios:
\begin{itemize} \itemsep0em
    \item \textbf{Scenario 1:} \acp{UE} are concentrated in a 100 m radius, representing a classic hotspot scenario;
    \item \textbf{Scenario 2:} \acp{UE} are less concentrated, distributed over a 500 m radius circle.
\end{itemize}

% start a new page without indent 4.6cm
%\clearpage
\end{paracol}
\nointerlineskip
\begin{specialtable}[H] 
\widetable 
\caption{\label{tab:icicparamPerformaceComparison} {Parameters used for the comparison of the ICIC algorithms in preliminary analysis A~\cite{baldo}.}}
\begin{tabular}[t]{p{6.5cm}p{6.5cm}}
\midrule
\textbf{Hard Frequency Reuse}   \\ 
\midrule
Bandwidth: cells 1 and 2       & 8 RBs           \\ 
\midrule
Bandwidth: cell 3              & 9 RBs           \\ 
\midrule
\vspace{0.05cm}
\textbf{Strict Frequency Reuse}  \\
\midrule
Bandwidth (center/edge)         & 6 RBs each         \\ 
\midrule
RsrqThreshold                   & 32                 \\ 
\midrule
Power offset (center)           & --6 dB              \\ 
\midrule
Power offset (edge)             & 3 dB               \\
\midrule
\vspace{0.05cm}
\textbf{Soft Frequency Reuse} \\
\midrule
Bandwidth (center)                   & 25 RBs     \\ 
\midrule
Bandwidth (edge): cells 1 and 2      & 9 RBs      \\ 
\midrule
Bandwidth (edge): cell 3             & 9 RBs      \\ 
\midrule
RsrqThreshold                        & 32         \\ 
\midrule
Power offset (center)                & --6 dB      \\ 
\midrule
Power offset (edge)                  & 3 dB       \\ 
\midrule
\vspace{0.05cm}
\textbf{Fractional Soft Frequency Reuse} \\
\midrule
Bandwidth (center/edge)             & 6 RBs each             \\ 
\midrule
RsrqThreshold (center)              & 37                     \\ 
\midrule
RsrqThreshold (edge)                & 32                     \\ 
\midrule
Power offset (center)               & --6 dB                  \\ 
\midrule
Power offset (middle)               & --1.77 dB               \\ 
\midrule
Power offset (edge)                 & 3 dB                   \\ 
\midrule
\end{tabular}
\end{specialtable}
\begin{paracol}{2}
%\linenumbers
\switchcolumn

\begin{specialtable}[H]
\caption{{Simulation parameters for preliminary analysis a: performance comparison of ICIC algorithms.}}
\label{tab:paramPerformaceComparison}

\begin{tabular}{m{6.5cm}m{6.5cm}}  
\toprule
\textbf{Parameter}              & \textbf{Value}\\
\midrule
Bandwidth (\acsp{RB})           & 25                       \\     
\midrule                                           
UE distribution                 & Uniform                   \\
\midrule
Total \acsp{UE}                 & 40 to 240                 \\
\midrule
Cell-edge \acsp{UE}             & 10 to 60                   \\
\midrule
Inter-\acsp{eNB} distance  & 1000 m                     \\
\midrule
Scheduling algorithm            & \textit{Proportional Fair}\\
\midrule
Channel model                   & Friis model            \\
\midrule
Error model                     & MIESM                    \\
\midrule
\acs{UE} mobility               & No mobility              \\
\midrule
Traffic model & Non-GBR TCP-based \\ & Video (Buffered Stream)   \\
\bottomrule
\end{tabular}
\end{specialtable}

% --------------------------------------------------------------------

\subsection{Results and Discussions}

For Scenario 1 (concentrated users), Figures \ref{fig:pcTPUTscenario1} and \ref{fig:pcSINRscenario1} present the 10th percentile throughput and the \ac{SINR} \ac{CDF}, respectively. The curves with dotted lines represent cell-edge users and the continuous represent cell-center users.

Regarding throughput, Figure \ref{fig:pcTPUTscenario1} shows that the \ac{NoOp} algorithm presents the best performance for cell-center users. All cells operate on the same frequency band; thus, more bandwidth is available, and a higher data rate is obtained. However, this is also the reason why \ac{NoOp} has the worst performance for edge users, {considering} these \acp{UE} are more affected by interference.

\begin{figure}[H]
\includegraphics[width=0.7\linewidth]{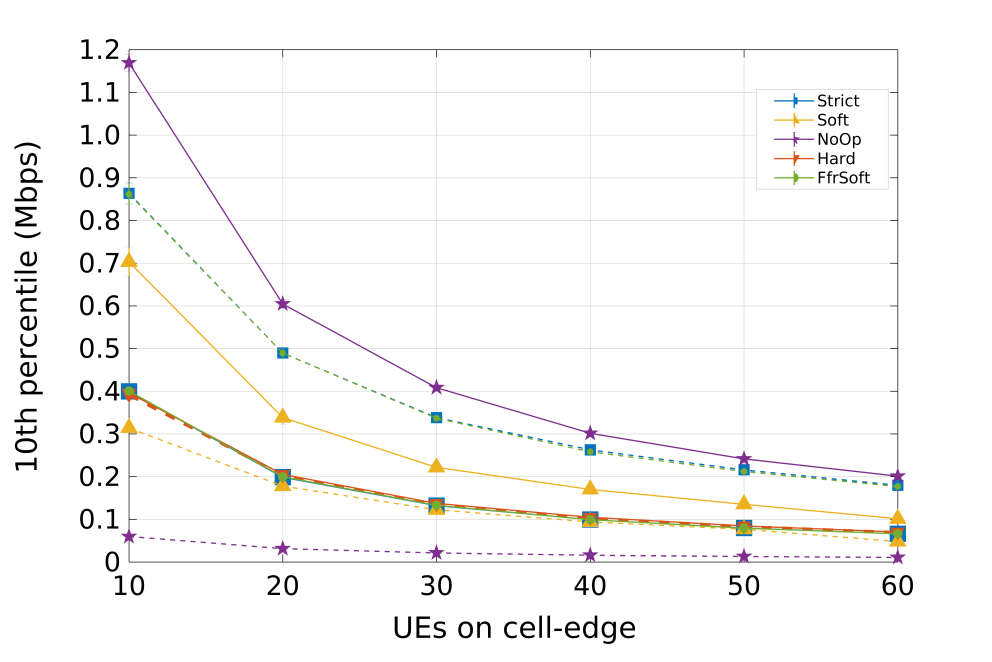}
\caption{\label{fig:pcTPUTscenario1} {Scenario 1 from preliminary analysis A: 10th percentile throughput for concentrated users (R = 100m). Each curve represents a different ICIC algorithm.}}
\vspace{0.5cm}

\end{figure}

\ac{Strict FR} and \ac{SFFR} have similar performance. They present the best results for cell-edge users, but limited performance for cell-center users. These algorithms allocate the same number of \acp{RB} for each region{. H}ence, cell-center users only have 25\% of the bandwidth available (Table \ref{tab:icicparamPerformaceComparison}). The \ac{HFR} algorithm has similar performance for both center and edge users, since it does not divide the cell into distinct regions. Users do not suffer from inter-cell interference, but the smaller bandwidth leads to lower throughput and both \ac{Strict FR} and \ac{SFFR} have better performance for all \acp{UE}.

The \ac{SFR} obtains good performance for cell-center users but poor performance for cell-edge users. The high number or \acp{RB} reserved for cell-edge users was not enough to balance \ac{ICI}, since the cell-edge bandwidth is shared with cell-center regions from adjacent cells.

Figure \ref{fig:pcSINRscenario1} indicates that cell-edge users have lower \ac{SINR} for all algorithms, and the worst results for both sets of users are from \ac{SFR} and \ac{NoOp}. \ac{HFR} has the best performance for cell-center users, but \ac{Strict FR} and \ac{SFFR} have better performance for cell-edge users while also maintaining good performance on cell-center, similar to \ac{HFR}. Moreover, considering the lower spectral efficiency, as discussed before, the \ac{HFR} algorithm is not able to provide good throughput levels.

\begin{figure}[H]
\includegraphics[width=0.7\linewidth]{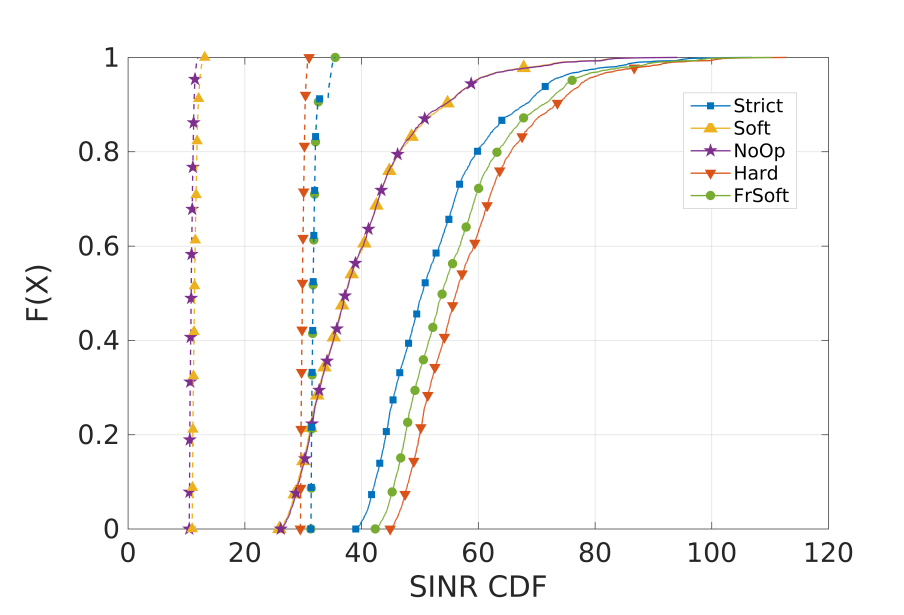}
\caption{\label{fig:pcSINRscenario1} {Scenario 1 from preliminary analysis A: SINR CDF for concentrated users (R = 100 m). Each curve represents a different ICIC algorithm.}}
\vspace{0.3cm}

\end{figure}

Figures \ref{fig:pcTPUTscenario2} and \ref{fig:pcSINRscenario2} show the results for Scenario 2, with less concentrated users. Concerning throughput (Figure \ref{fig:pcTPUTscenario2}), there is a significant performance loss for \ac{NoOp} and \ac{SFR}, mainly because cell-center \acp{UE} can interfere with cell-edge \acp{UE}. As a result, spreading the users can increase interference, due to higher occupation of the cell-center region. Additionally, \ac{Strict FR} and \ac{SFFR} still have the best throughput levels for cell-edge users.

\begin{figure}[H]
\includegraphics[width=0.7\linewidth]{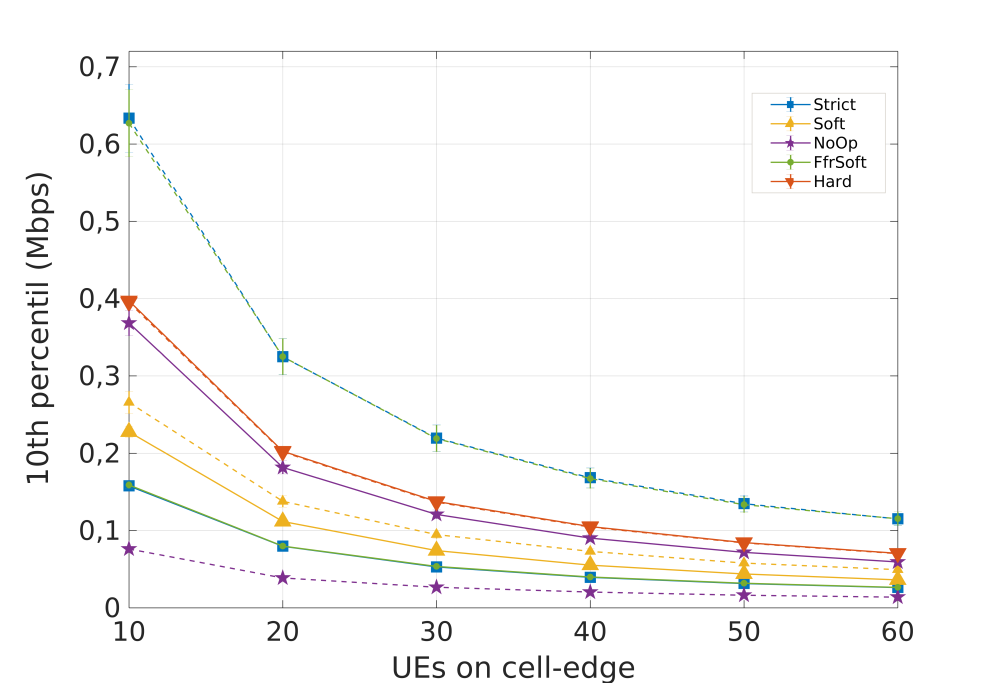}
\caption{\label{fig:pcTPUTscenario2} {Scenario 2 %please change hyphen (-) to minus (−).
 from preliminary analysis A: 10th percentile throughput for less concentrated users (R = 500 m). Each curve represents a different ICIC algorithm.}}
\vspace{0.3cm}

\end{figure}

Regarding \ac{SINR} results, Figure \ref{fig:pcSINRscenario2} shows that \ac{NoOp} and \ac{SFR} still have the worst performance in all cases and the \ac{Strict FR} and \ac{SFFR} algorithms maintain good performance on cell-edge. Nevertheless, all algorithms present lower \ac{SINR} on cell-edge, since, in this case, spreading the users led to more populated cell-edge regions. 

\begin{figure}[H]
\includegraphics[width=0.7\linewidth]{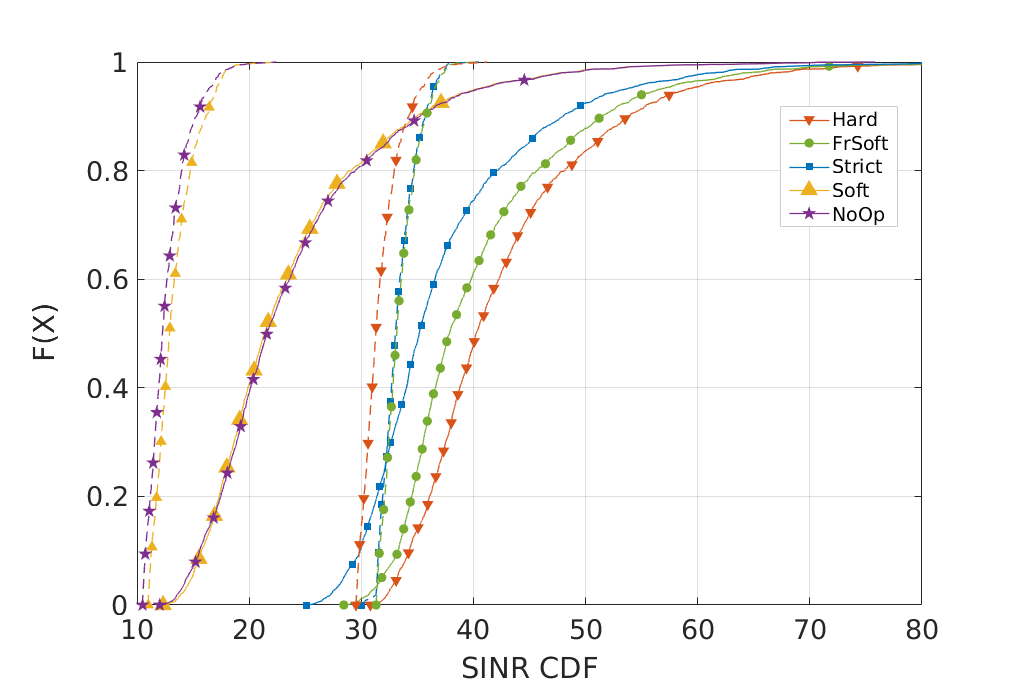}
\caption{\label{fig:pcSINRscenario2} {Scenario 2 from preliminary analysis A: SINR CDF for less concentrated users (R = 500 m). Each curve represents a different ICIC algorithm.}}
\vspace{0.3cm}

\end{figure}

Some important conclusions can be outlined:

\begin{itemize} \itemsep0em

\item \acs{FFR}-based \ac{ICIC} techniques can improve the performance of a mobile network, given the high interference suffered by its users. It is possible to {enhance} cell-edge performance without compromising cell-center users;

\item Simply changing the user density can strongly impact on system performance, {especially} if the bandwidth is divided {into} sub-bands. The concentration or spreading of the users can result in different occupation of the sub-bands, leading to overloaded or nearly empty sub-bands;

\item There is no scheme that performs best in every situation. However, the \ac{Strict FR} and the \ac{SFFR} schemes have a good performance in different scenarios, and they are both efficient at reducing \ac{ICI}, {especially} for cell-edge users;

\item The \acf{Strict FR} algorithm has the best compromise between performance and complexity, {considering} \ac{SFFR} has a higher number of sub-bands, which leads to more complexity when adjusting the parameters. Therefore, the \ac{Strict FR} is the object of the study presented in the following sections.

\end{itemize}

\section{{Preliminary Analysis B: Factorial Design using the Strict Frequency \mbox{Reuse Algorithm}}}\label{sec:2kfactorial}

The second step towards the proposed solution consists {of} selecting one of the \ac{ICIC} algorithms to evaluate the impact of its most relevant parameters on system performance. This analysis has two further steps: a 2k factorial design and a full factorial design.

These studies will be conducted in the \ac{Strict FR} algorithm for its good compromise between complexity and performance.

\subsection{Evaluation Scenario}

The proposed scenario is very similar to the one described in Section \ref{sec:scenarioPerformanceComparison}, as illustrated in Figure \ref{fig:cenario2k}. It consists {of} three \acp{eNB} positioned in the vertices of an equilateral triangle with side equal to 1000 m. A total of 80 users are randomly positioned in the entire scenario according to {a} uniform distribution. They do not have mobility and each \ac{UE} is served by the closest \ac{eNB}. The simulation parameters are presented in Table \ref{tab:paramParametric}.

\begin{figure}[H]
\includegraphics[width=0.4\linewidth]{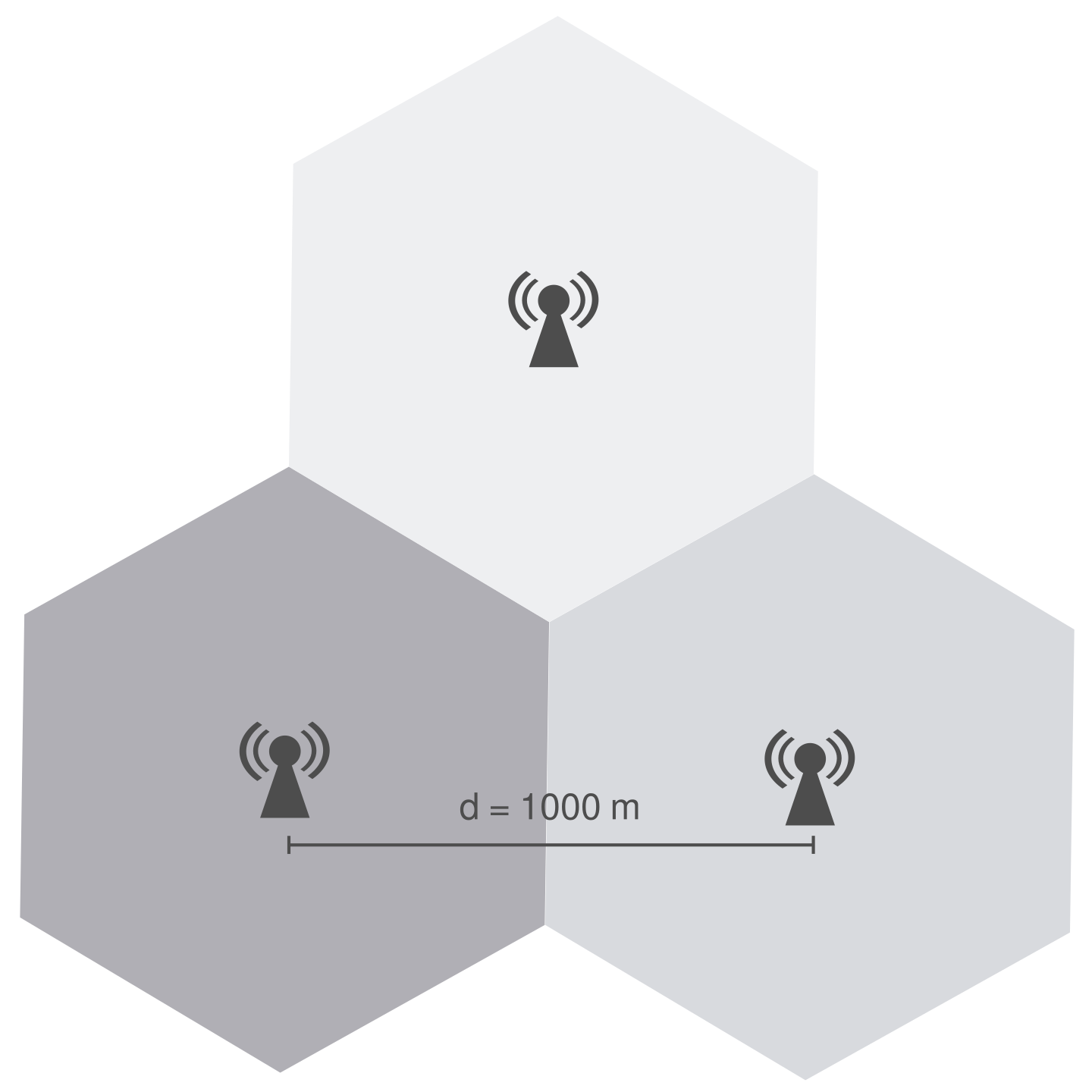}
\caption{\label{fig:cenario2k} {Scenario for preliminary analysis B, which is similar to the scenario in Figure \ref{fig:scenarioPerformanceComparison}, but without the radius R, since all users are randomly distributed according to a uniform distribution.}}
%\vspace{0.45cm}

\end{figure}
\vspace{-6pt}

\begin{specialtable}[H]
\caption{{Simulation parameters for preliminary analysis B: parametric analysis of strict \mbox{frequency reuse.}}}
\label{tab:paramParametric}

\begin{tabular}{m{6.5cm}m{6.5cm}}  

\toprule
\textbf{Parameter}              & \textbf{Value}            \\
\midrule
Bandwidth (\acsp{RB})           & 25                        \\     
\midrule                                          
UE distribution                 & Uniform                   \\
\midrule
Number of \acsp{UE}             & 80                        \\
\midrule
Distance between \acsp{eNB} (m) & 1000                      \\
\midrule
Scheduling algorithm            & \textit{Proportional Fair}\\
\midrule
Channel model                   & Friss model \\
\midrule
Error model                     & MIESM                     \\
\midrule
\acs{UE} mobility               & No mobility               \\
\midrule
Traffic model                   & Non-GBR TCP-based         \\
                                & Video (Buffered Stream)   \\
\bottomrule

\end{tabular}
\end{specialtable}

\subsection{$2^k$ Factorial Design}

The $2^k$ factorial design (2k factor) is a specific case of factorial design that identifies which parameters of an experiment have {a} significant effect on {the} desired output \cite{montgomery}. It provides a better understanding of the system, and can simplify future analysis. For example, assume that the 2k factor is conducted for an experiment {with} a parameter called $P$. If the 2k factor determines that $P$ is not relevant, a following full factorial design will need fewer repetitions, since $P$ does not need to change. %Footnote is not permitted in this journal, so we have moved it into the text, please confirm the whole text.
 {A Full Factorial Design consists of replicating an experiment for all possible combinations of parameters}.

To perform the 2k factor, $k$ factors (parameters) are chosen, and they assume only two distinct values, preferably a low and a high value close to the upper and lower limits of the parameter's range. Besides, it is usually followed by an \acf{ANOVA} that validates the results by statistically rejecting or not the null hypothesis, which is a given parameter does not affect the desired output. \ac{ANOVA} tests this hypothesis by comparing variances. 

Consider the value $F0$ in Equation \eqref{eq:f0}. This value is calculated for each parameter, and it is defined as the division between two \acfp{SS}, which is a non-biased estimator of a population's variance \cite{montgomery}. For example, if a simulation that has a parameter A is repeated $n$ times for each value of the parameter, $SS_{treat}$ is the \ac{SS} that measures the variation between the different values of parameter A and $SS_{error}$ is the \ac{SS} that measures the variation between the $n$ results of a single value of A. If the change of a given parameter has {a} relevant impact on the desired output, the null hypothesis is rejected, and $SS_{treat}$ is bigger than $SS_{error}$ \cite{montgomery}. Therefore, the bigger $F0$ is, the bigger is the parameter's impact.

\begin{equation}\label{eq:f0}
F0 = \frac{SS_{treat}}{SS_{error}}
\end{equation}

However, it is recommended to define a threshold for $F0$ that guarantees the parameter's relevance. Since $SS_{treat}$ and $SS_{error}$ are both chi-square random variables by \mbox{definition \cite{chisquare},} $F0$ has a Fisher--Snedecor distribution \cite{montgomery,fdistribution}. Defining a level of significance $\alpha = 0.001$, for the evaluated parameters, the threshold value for $F0$ is 10.83. Thus, $F0$ is calculated for each parameter. If $F0$ is greater than 10.83, the parameter has {a} significant impact on the output.

Therefore, the $2^k$ factorial was conducted for the \ac{Strict FR}'s parameters: \textit{CenterPowerOffset}, \textit{RsrqThreshold}, and \textit{BandwidthDistribution}. As detailed {in} Section \ref{sec:systemModel}, they define the power level of the common sub-band, the threshold for allocating users on sub-bands and the number of \acp{RB} on each sub-band, respectively. {In order to investigate all users and users with the worst performance (most likely cell-edge users), the targeted output is the average throughput and its 10th percentile.}

Tables \ref{tab:F0avtp} and \ref{tab:F010percentil} present the calculated $F0$ for each parameter and for the interaction between parameters. The nomenclature X * Y represents the interaction between parameters X and Y. CenterPowerOffset is the only parameter that does not have {a} significant impact on the average or 10th percentile throughput. This is due to how bandwidth is allocated in the \ac{Strict FR}. The cell center of each cell does not share bandwidth with {the} cell-edge of any cell. Hence, users from different cells allocated to the same bandwidth might not be close enough to cause significant \ac{ICI}.

\begin{specialtable}[H]
\caption{F0 results for average throughput.}
\label{tab:F0avtp}

\begin{tabular}{m{8.5cm}m{4.5cm}}  

\toprule
\textbf{Parameter (Average Throughput)}     & \textbf{F0}   \\
\midrule
CenterPowerOffset (A)			            & 0.084         \\
\midrule
RsrqThreshold (B)                           & 39040.4       \\
\midrule
BandwidthDistribution (C)                   & 4301.4        \\
\midrule
BandwidthDistribution * CenterPowerOffset   & 0.0015        \\
\midrule
BandwidthDistribution * RsrqThreshold       & 6264.7        \\
\midrule 
RsrqThreshold * CenterPowerOffset           & 0.0012        \\
\midrule
A * B * C                                   & 0.00032       \\
\bottomrule

\end{tabular}
\end{specialtable}

\vspace{-6pt}
\begin{specialtable}[H]
\caption{F0 results for the 10th percentile.}
\label{tab:F010percentil}

\begin{tabular}{m{8.5cm}m{4.5cm}}  

\toprule
\textbf{Parameter (10th Percentile)}        & \textbf{F0}       \\
\midrule
CenterPowerOffset (A)		                & 0.00059           \\
\midrule
RsrqThreshold (B)                           & 2688.37           \\
\midrule
BandwidthDistribution (C)                   & 6359.40           \\
\midrule
BandwidthDistribution * CenterPowerOffset   & 0.035             \\
\midrule
BandwidthDistribution * RsrqThreshold       & 695.91            \\
\midrule 
RsrqThreshold * CenterPowerOffset           & 0.0033            \\
\midrule
A * B * C                                   & 0.00034           \\
\bottomrule

\end{tabular}
\end{specialtable}

\subsection{Full Factorial Design}\label{sec:fullFactorial}

After identifying which parameters are relevant, a full factorial design repeats the simulation to vary these parameters. The purpose of this investigation is to evaluate how system performance is affected by the \ac{Strict FR}'s parameters, in order to identify the best configurations. {Preliminary tests indicated that 100 repetitions are enough for the proposed scenario to provide a good confidence interval.} Moreover, given that CenterPowerOffset represents a power level and does not impact on throughput, it is fixed on its minimum possible value, lowering energy consumption. Each evaluated parameter is detailed in Table \ref{tab:fullFactorial}.

\begin{specialtable}[H]
\caption{Full factorial design: parameter configurations.}
\label{tab:fullFactorial}

\begin{tabular}{m{6.5cm}m{6.5cm}}  

\toprule
\textbf{Parameter} & \textbf{Values} \\
\midrule
RsrqThreshold & 24, 25, 26, 27, 28, 29, 30, 31, 32, 33, 34 \\    
\midrule                                      
Common/Private & Bandwidth Distribution 1: 6 / 18  \\
Sub-band (RBs) & Bandwidth Distribution 2: 12 / 12 \\
               & Bandwidth Distribution 3: 18 / 6 \\
\bottomrule

\end{tabular}
\end{specialtable}

Figures \ref{fig:2kAvgTput} and \ref{fig:2k10thTput} present the results for the average and 10th percentile throughput, respectively. Both have a 95\% confidence interval. Lower values of RsrqThreshold result in better performance if more bandwidth is allocated to cell-center users. Consequently, bandwidth distribution 1, {which} allocates 6~\acp{RB} for cell-center and 18~\acp{RB} for cell-edge users (6~\acp{RB} for each cell edge), has the worst performance. This is due to more users being allocated to the cell center as RsrqThreshold gets lower, causing the common sub-band \mbox{to overload.}

\begin{figure}[H]
\includegraphics[width=0.7\linewidth]{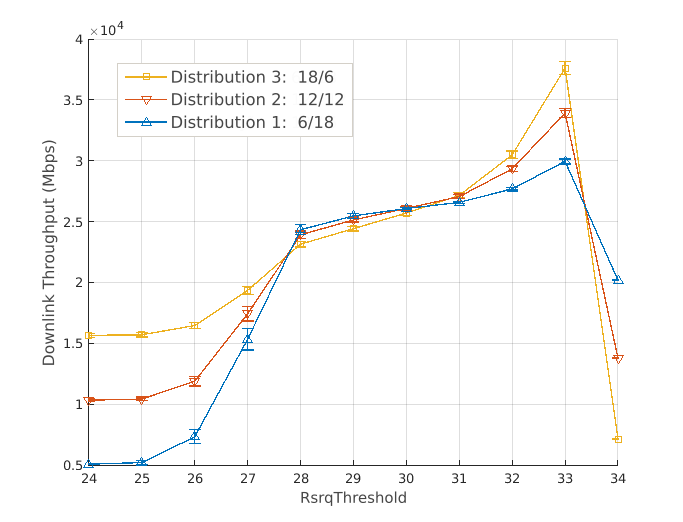}
\caption{\label{fig:2kAvgTput} {Full factorial design using the strict frequency reuse: results for the average throughput. Each curve is a different bandwidth distribution.}}
%\vspace{0.5cm}

\end{figure}

Moreover, increasing the RsrqThreshold yields higher throughput for all distributions (Figure \ref{fig:2kAvgTput}) until each curve reaches its peak. At this point, the distribution with more band allocated to the center still has the best performance. From that value, the private sub-band gets overloaded, which results {in} performance loss. This loss is more severe for distributions with less bandwidth allocated to the edge.

Figure \ref{fig:2k10thTput} has similar behavior to Figure \ref{fig:2kAvgTput}, since increasing the RsrqThreshold results in higher throughput. However, each curve starts decreasing at different values. Distributions with less bandwidth to the edge sub-band have {their} peak at lower RsrqThreshold values. At RsrqThreshold~=~33, Figure~\ref{fig:2kAvgTput} indicates that throughput is better for distributions with more bandwidth allocated to the center. Figure~\ref{fig:2k10thTput} indicates the opposite, and bandwidth distributions 1 and 2 have similar performance on RsrqThreshold~=~32.

\begin{figure}[H]
\includegraphics[width=0.7\linewidth]{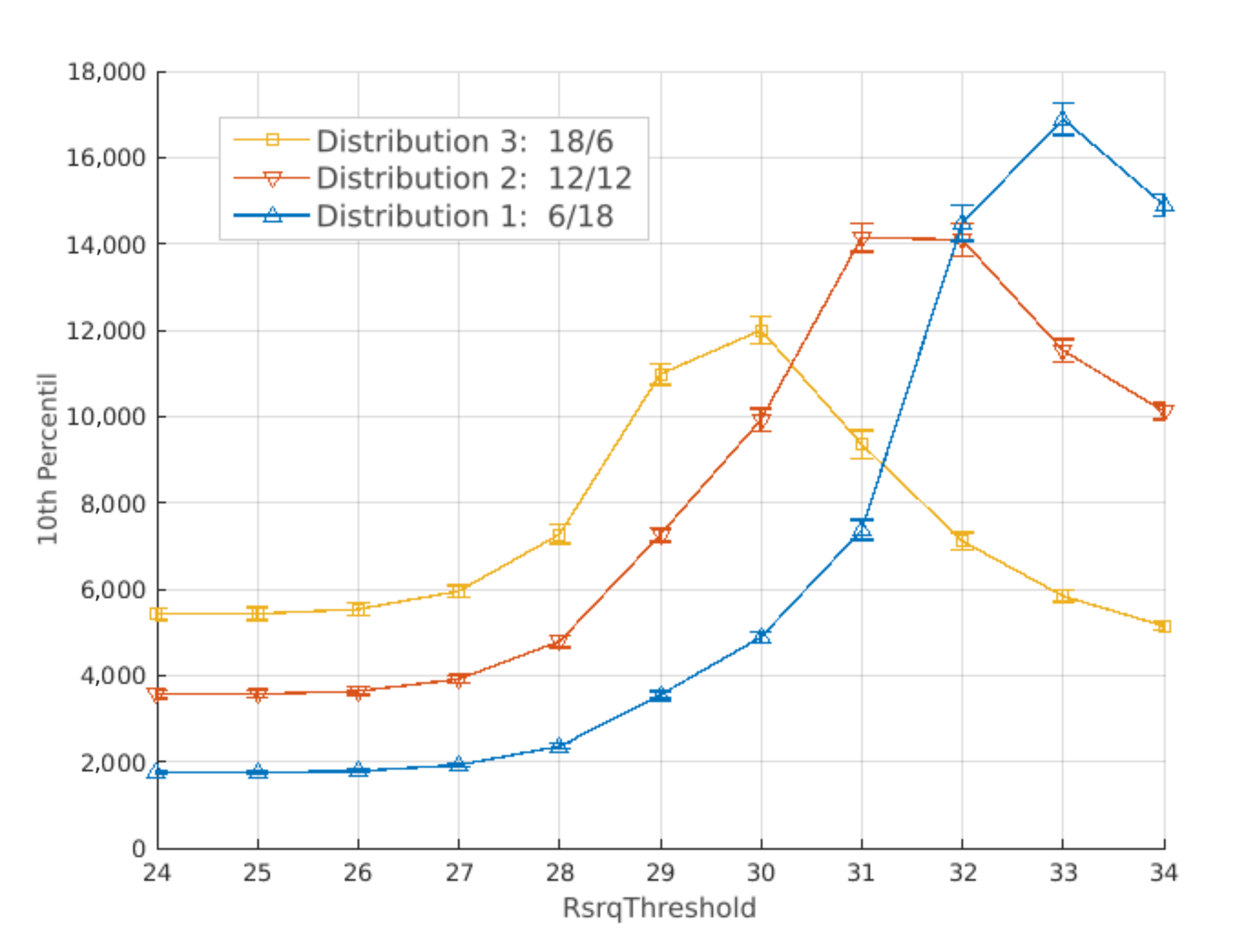}
\caption{\label{fig:2k10thTput} {Full %Please add a comma for more than 4-digit numbers. For example, please change "2000" to "2,000".
 factorial design using the strict frequency reuse: results for the 10th percentile throughput. Each curve is a different bandwidth distribution.}}
%\vspace{0.5cm}

\end{figure}

Lastly, Figure~\ref{fig:2kSinr} presents the \ac{SINR} \ac{CDF} for RsrqThreshold~= ~3. At this value, more bandwidth allocated for cell-edge users result in better \ac{SINR}.

\begin{figure}[H]
\includegraphics[width=0.7\linewidth]{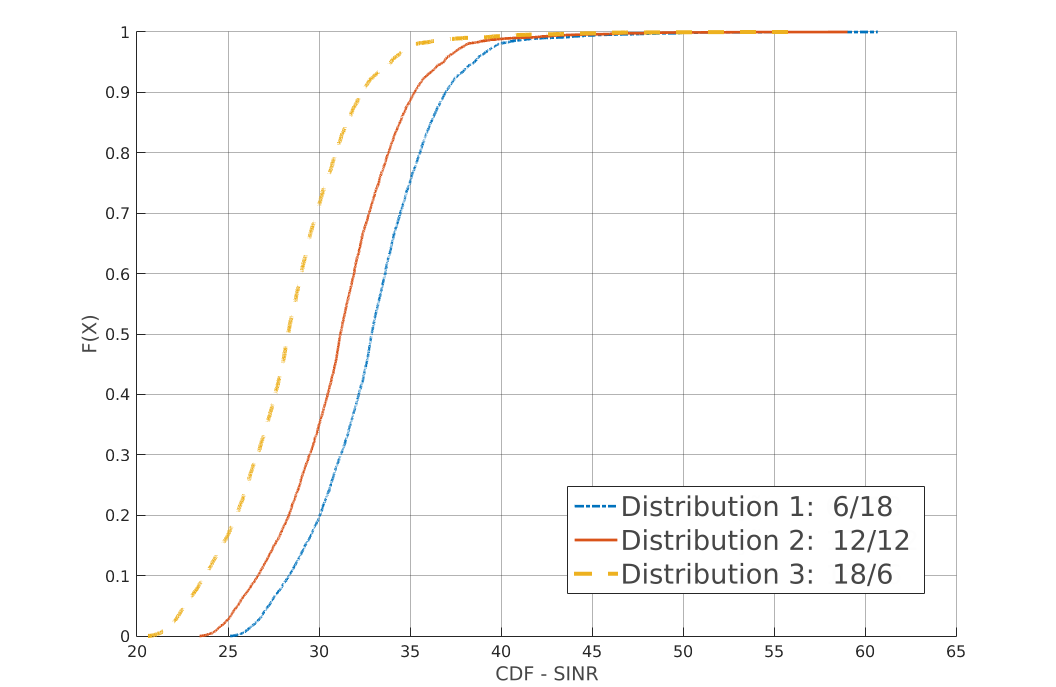}
\caption{\label{fig:2kSinr} {Full factorial design using the strict frequency reuse: results for the SINR CDF. Each curve is a different bandwidth distribution.}}
%\vspace{0.5cm}

\end{figure}

These results show a clear compromise between the average user performance and the performance of cell-edge users. Besides, in a scenario that hotspots appear unpredictably, certain regions can also become overloaded. Since the RsrqThreshold has direct impact in user allocation, its dynamic variation can efficiently mitigate the performance loss caused by the hotspots.

%------------------------------------------------------------------

\section{{Preliminary Analysis C: Evaluating the Hotspot Scenario}} \label{sec:hotspot}

The last set of investigations before presenting the proposed
solution evaluates the impact of hotspots on system performance. The goal is to compare a scenario with hotspots to a scenario with users distributed uniformly.

\subsection{Evaluation Scenario}

The evaluation scenarios are similar to those in Sections
\ref{sec:scenarioPerformanceComparison} and \ref{sec:fullFactorial}, since there are three \acp{eNB} equally distanced by 1000~meters. In this analysis, scenarios 1 and 2 have uniformly distributed users, while scenario 3 also has 5 hotspots with 15 users each. {Figure \ref{fig:hotspotScenario} illustrates the approximate position of each hotspot, which are not random.} The differences between these scenarios are summarized in Table \ref{tab:scenariosHotspot}.

There are two scenarios without hotspots, but with different amounts of users. As a result, it is possible to evaluate the impact of hotspots that increase (or not) the total amount of users in the system. Other simulation parameters are presented in Table~\ref{tab:paramHotspot}.

\begin{figure}[H]
\includegraphics[width=0.3\linewidth]{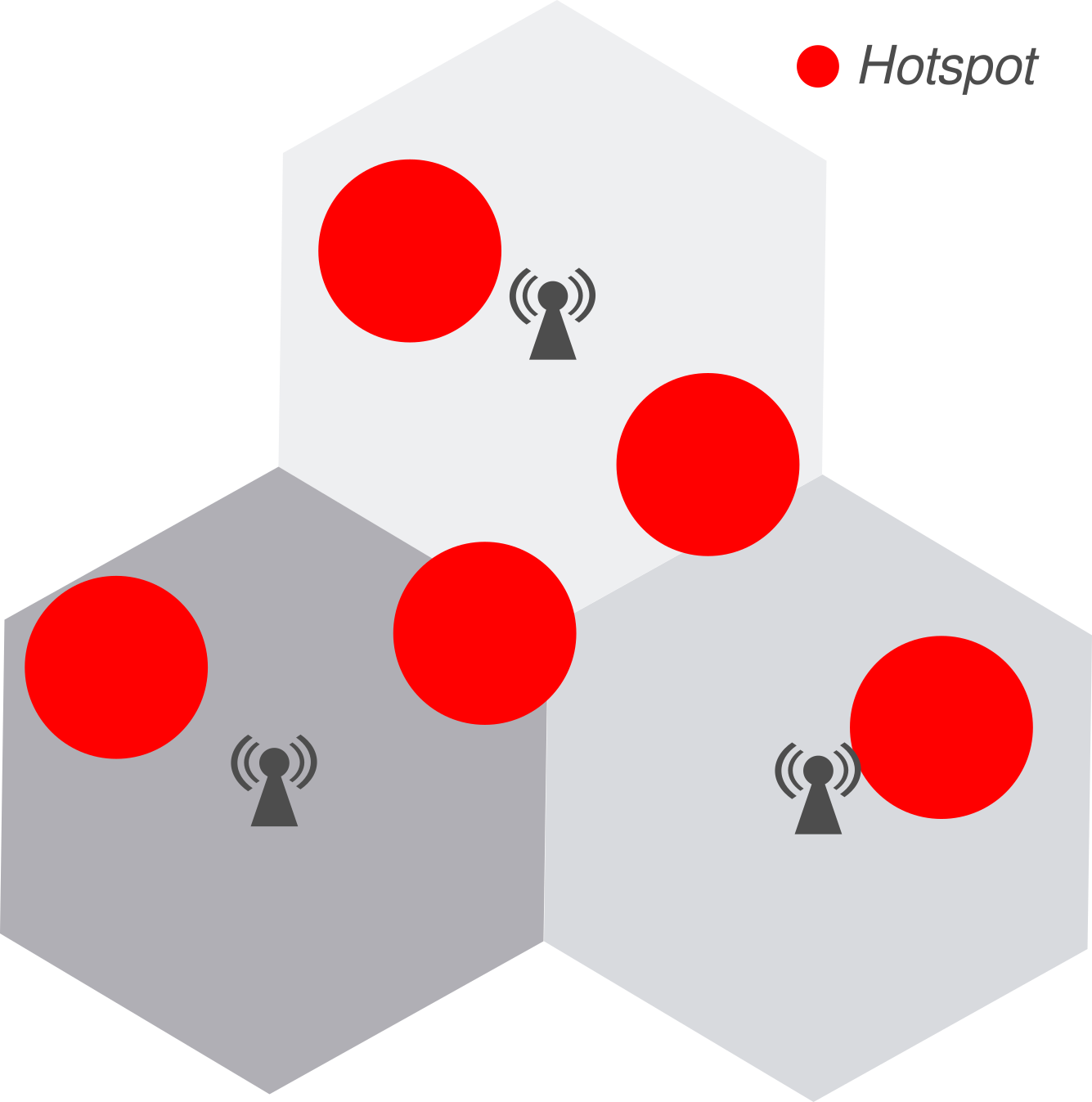}
\caption{\label{fig:hotspotScenario} {Scenario for preliminary analysis C with the approximate position for each hotspot, which are not randomly located.}}
\vspace{0.35cm}

\end{figure}

\begin{specialtable}[H]
\caption{Difference between hotspot scenarios.}
\label{tab:scenariosHotspot}

\begin{tabular}{m{13cm}}  

\toprule
\textbf{Scenario 1}                 \\
\midrule                                             
60 uniformly distributed users and no hotspots     \\
\midrule
\textbf{Scenario 2}                 \\
\midrule
135 uniformly distributed users and no hotspots    \\
\midrule
\textbf{Scenario 3}                 \\
\midrule
135 users: 60 uniformly distributed users and   \\    
5 hotspots with 15 users each       \\
\bottomrule

\end{tabular}
\end{specialtable}
\vspace{-6pt}

\begin{specialtable}[H]
\caption{Simulation parameters: hotspot scenarios.}
\label{tab:paramHotspot}

\begin{tabular}{m{6.5cm}m{6.5cm}}  

\toprule
\textbf{Parameter}              & \textbf{Value}            \\
\midrule
Bandwidth (\acsp{RB})           & 100                       \\     
\midrule                                             
UE distribution                 & Uniform / Hotspots        \\
\midrule
Number of \acsp{UE}             & 60 / 135                 \\
\midrule
Distance between \acsp{eNB} (m) & 1000                      \\
\midrule
Scheduling algorithm            & \textit{Proportional Fair}\\
\midrule
Simulation duration             & 6000 subframes            \\
\midrule
Channel model                   & Friis Model                \\
\midrule
Error model                     & MIESM                     \\
\midrule
\acs{UE} mobility               & No mobility               \\
\midrule
Traffic model                   & Non-GBR TCP-based         \\
                                & Video (Buffered Stream)   \\
\bottomrule

\end{tabular}
\end{specialtable}

\subsection{Simulation Results}

Figure \ref{fig:hotspotTput} presents the simulation results in terms of the 10th percentile of throughput. The dashed lines represent the scenario with hotspots (Scenario 3). The simulations were executed using the \ac{Strict FR} algorithm, with the variation of its relevant parameters, according to Section \ref{sec:2kfactorial} (RsrqThreshold and BandwidthDistribution). The former ranges between 30 and 33, and the latter is evaluated in two configurations. The first allocates 52 \acp{RB} to the common sub-band and 16 \acp{RB} to the private sub-band. The second allocates 28 \acp{RB} to the common sub-band and 24 \acp{RB} to the private sub-band.

\begin{figure}[H]
\includegraphics[width=0.7\linewidth]{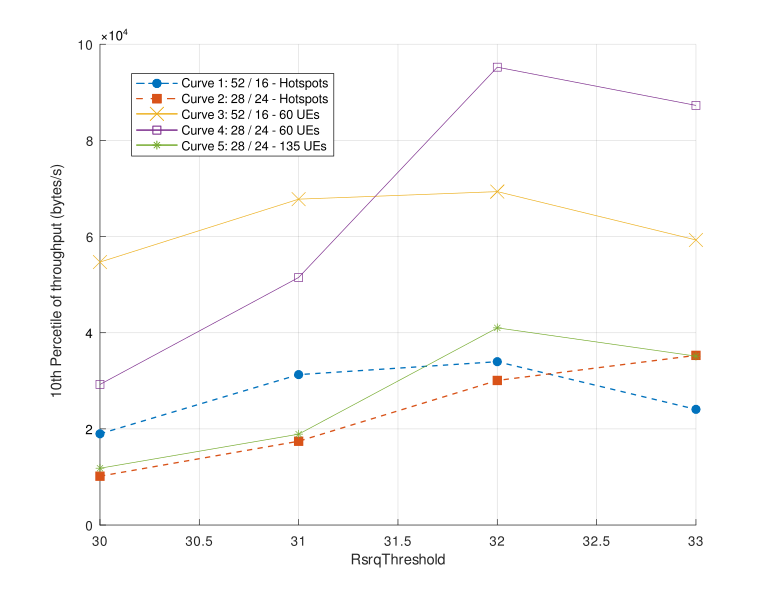}
\caption{\label{fig:hotspotTput} {Simulation results in terms of the 10th percentile of throughput. The dashed lines represent the scenario with hotspots.}}
%\vspace{0.3cm}

\end{figure}

The 10th percentile of throughput (Figure \ref{fig:hotspotTput}) indicates a significant performance loss due to the hotspots. The curves {with} the same distribution (curves 2, 4, and 5) show that the scenario with hotspots has lower throughput, {especially} if compared to curve 4, {which} has fewer users.

Even though the performance loss is expected, given the increase in the {number} of users, both cases discussed above are important. Section \ref{sec:interferenceHotspot} discussed scenarios naturally characterized by hotspots. In most of these scenarios, the hotspots also increase the total number of active users in the system, such as football games or music festivals. However, even if the change is only in user concentration, there is a relevant performance loss, as presented {in} Figure \ref{fig:hotspotTput}, given that curve 2 has {the} worst performance than curve 5.

%-------------------------------------------------------------------

\section{{Proposed Solution: Background, Implementation, and Simulation Results}}\label{sec:solution}

The results presented in Section \ref{sec:performanceComparison} show that \acs{FFR}-based techniques efficiently mitigate \ac{ICI}, improving throughput and \ac{SINR}. It also shows that no algorithm performs best in every situation, but the \ac{Strict FR} has a good compromise between performance and complexity. Moreover, not all of its parameters have {a} significant impact on throughput and \ac{SINR}, according to Section \ref{sec:2kfactorial}.

Results from Section \ref{sec:hotspot} indicate that hotspots can have {a} significant impact on system performance, and different states of the system may require different configuration{s} of the \ac{Strict FR} parameters to achieve %optimal 
improved performance. Furthermore, Section \ref{sec:interferenceHotspot} discusses challenges related to hotspot scenarios and highly dynamic urban areas. In such scenarios, techniques that allow the system to adapt dynamically may increase performance.

Therefore, this section presents our proposed solution for scenarios where hotspots appear unexpectedly. This solution applies \ac{ML} techniques to dynamically regulate the \textit{RsrqThreshold} of the Strict~FR.

\subsection{Machine Learning Techniques}

\acf{ML} is a subset of artificial intelligence that builds mathematical models to find patterns in data sets. The goal is to make decisions without human assistance. They have been widely discussed in the literature {for their ability to adapt to} changing scenarios and reduce site planning. For instance, the authors of \cite{TrejoNarvaez2018} present various \ac{ML} algorithms and how they have been employed to coordinate co-channel interference. 

Moreover, these techniques can be classified as supervised, non-supervised, and \acf{RL}. Supervised learning algorithms are akin to a supervisor that holds the desired knowledge. These algorithms adapt based on inputs previously known by the supervisor (labeled data sets). Non-supervised algorithms do not have labeled inputs{. H}ence, they aim to find hidden patterns or similarities within the data. %\ac{RL}, on the other hand, 
On the other hand, \ac{RL} continuously interacts with the system, and each decision produces a reward or punishment that serves to update the algorithm parameters. The goal is to maximize or minimize the reward or punishment \cite{haykin2009neural}.

For the proposed scenario, \ac{RL} is the most appropriate paradigm. Supervised learning requires prior knowledge and dedicated time for training, which is not desirable for a system that demands constant re-adapting. Non-supervised learning does not require prior knowledge, but \ac{RL} is a better option for this task, {considering} the algorithm learns from each action taken {using} immediate rewards. \ac{RL} is an efficient paradigm for interaction with uncertain environments \cite{livro-RL}, enabling real-time decision making.

\subsection{Reinforcement Learning}

\acf{RL} is a \ac{ML} paradigm that has a learning agent that attempts to reach an objective through trial and error, i.e., through constant interactions with the system. The agent is not instructed on which action should be taken, but it must be capable {of observing} the state of the system and take actions that can change it.

As a consequence of the action, the system provides a numerical reward and informs its new current state. The agent must identify which actions result {in} bigger rewards in any state of the system. Hence, the algorithm estimates the reward for each action $a_n$ for each state $s_n$. However, the estimated value is not based on the current action alone. It considers all of the actions taken so far \cite{livro-RL}.

Among \ac{RL} approaches, \acf{QL} has been constantly discussed in the literature, {especially} because it does not require {estimating} the dynamics of the environment (it is model-free)~\cite{TrejoNarvaez2018}.

For example, the authors of \cite{Chen2017} attempt to maximize system performance in an ultra-dense heterogeneous scenario. They propose a dynamic resource allocation scheme using centralized and distributed \ac{QL}. The authors of \cite{Simsek2012} use \ac{QL} in a \ac{HetNet} scenario. Each cell learns optimum values for the \ac{CRB} and the \ac{DL} power transmission level. The results show a 125\% improvement compared to static \ac{ICIC} techniques. Additionally, the authors of \cite{Morozs2016} evaluate a scenario where a stadium is served by 78 \acp{eNB}. They propose an improved \ac{QL} algorithm that decreases the convergence time. 

Our previous works include \cite{gppcom-dm-csat,gppcom-gtdm-csat}. They present solutions that use \ac{ABS} and \ac{QL} to coordinate the shared transmission between \ac{LTE} and Wi-Fi in the 5~GHz bandwidth. Therefore, this paper applies the methodology previously tested, {especially} regarding the prototyping of a dynamic solution of \ac{RRM} using \ac{QL}.

%The authors of this paper are members of \ac{GppCom}, a research group that has expertise on dynamic algorithms using \ac{QL} and ns-3 prototyping. Previous work from other members of the group include \cite{gppcom-dm-csat} and \cite{gppcom-gtdm-csat}. They present solutions that use \ac{ABS} and \ac{QL} to coordinate the shared transmission between \ac{LTE} and WiFi in the 5~GHz bandwidth. Therefore, this paper applies the methodology previously tested, especially regarding the prototyping of a dynamic solution of \ac{RRM} using \ac{QL}.

\subsection{Q-Learning}\label{sec:q-learning}

\acf{QL} \cite{q-learning} operates by creating a table of q-values using a $Q(s,a)$ function that estimates the rewards for each state/action pair. $Q(s,a)$ is the q-value of action $a$ when the system state is $s$. The q-value indicates the desirability of taking action $a$ when in state $s$. The higher the q-value, the more desirable it is. 

The values estimated by this function are updated with each iteration, according to Equation~\eqref{eq:qlearning} \cite{livro-RL}: 

\begin{equation}\label{eq:qlearning}
Q(S_t,A_t) \leftarrow Q(S_t,A_t) + \alpha [ R_{t+1} + \gamma \max_a{Q(S_{t+1},a) -  Q(S_t,A_t)} ],
\end{equation}

\noindent where $R_{t+1}$ is the reward from action $A_t$ when the state is $S_t$; $\alpha$ is the learning rate, which determines the impact of new information in the update of the q-values; and $\gamma$ is a discount factor, which determines the impact of future rewards. Its value range{s} from 0 to 1, and values close to zero indicate that future rewards are not important. In this case, immediate rewards have {a} greater impact in the learning process. 

Therefore, the algorithm's goal is to find the optimum $Q(s,a)$ table, which gives an optimum policy $\pi^*$. This policy corresponds to choosing action $a$, given a state $s$, as defined in Equation~\eqref{eq:estrategia-qlearning}:

\begin{equation}\label{eq:estrategia-qlearning}
\pi^*(s) = arg\max_a Q(s,a).
\end{equation}

The author of \cite{phd_q_learning} demonstrates that, given enough samples, the \ac{QL} algorithm converges to a function directly derived from the Bellman equations, which is the classic solution of a \ac{MDP}. The key to that convergence is a Markovian Process named \ac{ARP}. It is possible to use the weak law of large numbers to prove that the policy $\pi$ gets closer to the optimum policy $\pi^*$ for a growing number of samples.

\subsection{{The Q-Learning Implementation}}

As a consequence of the discussions and exploratory analysis presented, the \ac{QL}-based solution aims at maximizing the average \ac{SINR} to mitigate the negative impact of \ac{ICI} on a dynamic scenario. The algorithm controls the \textit{RsrqThreshold}. {According to the results shown in Section \ref{sec:2kfactorial}, the \textit{RsrqThreshold} is the parameter that has the biggest impact on average throughput. Hence, it was the first option for the \ac{QL} algorithm. However, future works include the control of both the RsrqThreshold and the BandwidthDistribution, which is also an important parameter for the \acs{ICIC} performance} parameter, which dictates user allocation in the sub-bands defined by the \ac{Strict FR}.

The \ac{QL} algorithm operates as a centralized solution, namely that every decision affects all cells in a cluster. The \ac{ICIC} techniques presented in Section \ref{sec:icic-techniques} split the bandwidth for clusters formed by three cells. Therefore, each change in the RsrqThreshold affects all three cells of the cluster. If each cell had an independent \ac{QL} algorithm, they would be competing, as each action would affect the whole cluster, based only on the data from one cell.

The \ac{QL} algorithm needs the following parameters:

\begin{itemize} \itemsep0em
    \item A set of available actions, $A = {a_1, a_2, ..., a_n}$;  
    \item A set of possible states of the system, $S = {s_1, s_2, ..., s_n}$;  
    \item A $Q(s,a)$ matrix to store the estimated rewards;  
    \item $\alpha$ and $\gamma$.  
\end{itemize}

The algorithm was configured with six states and six actions. The metric that defines the state is the \ac{SINR}; however, this metric may vary. For example, depending on the service targeted, it could be throughput, \ac{PLR}, delay, or a combination of those, as presented in Equation~\eqref{eq:summation_metrics}~\cite{patent_vicente}:

\begin{equation}\label{eq:summation_metrics}
M_a = \sum_{i=1}^{n} w_i f_i (M_i),
\end{equation}

\noindent where $w_1, \; ..., \; w_n$ are the weights of each metric ($M_1, \; ..., \; M_n$) and $f_i$ is a normalization function that maps the metrics to values between 0 and 1.

The \ac{SINR} was chosen because it directly indicates the impact of interference in users. Hence, the average \ac{SINR} defines the states of the system and it is used as the reward from each state/action pair. Through experimental analysis to characterize the scenarios, the states are defined as follows:

\begin{itemize} \itemsep0em
    \item State 1: $ SINR < 24~dB $;
    \item State 2: $ 24~dB < SINR \leq 27~dB $;
    \item State 3: $ 27~dB < SINR \leq 29~dB $;
    \item State 4: $ 29~dB < SINR \leq 30~dB $;
    \item State 5: $ 30~dB < SINR \leq 32~dB$;
    \item State 6: $ SINR > 32~dB $.
\end{itemize}

Additionally, the set of available actions corresponds to RsrqThreshold values: \linebreak~$ A~=~\{29, 30, 31, 32, 33, 34\}$. These values were selected based on the results presented {in} Section~\ref{sec:2kfactorial}, given that this parameter has {a} bigger impact on system performance when operating on that range. 
The matrix $Q(s,a)$ initially has all values equal to zero and the values for $\alpha$ and $\gamma$ are $0.4$ and $0.5$, respectively. This choice is based on exploratory simulations and related works \cite{gppcom-dm-csat} {and the $Q(s,a)$ values are calculated according to Equation \ref{eq:q_values_pseudocode}:}

{
\begin{equation}\label{eq:q_values_pseudocode}
Q(s,a) \leftarrow (1-\alpha)Q(s,a) + \alpha\left [ r + \gamma \max\limits_{a}Q(s{'},a) \right ]
\end{equation}
}

Algorithm \ref{alg:pseudo_ql} presents a pseudo-code of the proposed solution. The \ac{QL} algorithm is executed each $40$~ms. This interval is inspired by the study presented in \cite{gppcom-dm-csat}; however, exploratory simulations were conducted to assure that the \ac{SINR} samples collected during this interval are sufficient to represent the average \ac{SINR} of the system.

As mentioned previously, the \ac{QL} seeks an optimum solution through a succession of actions. Thus, at any given time, the algorithm may get stuck on a set of state/action pairs without exploring other options (similar to an optimization algorithm that gets stuck on a local minimum). To address this issue, the parameter $\epsilon$ may be used to control the algorithm's level of exploration, making the agent choose a random action eventually.

\subsection{Evaluation Scenario}
\label{sec:scenario_ql}

The scenarios are similar to that of Section~\ref{sec:hotspot}. There are three \acp{eNB} equally distanced by 1000 m, and the users are allocated in two manners. The first group of users is randomly allocated in the entire scenario, using a uniform distribution. The second group is allocated on hotspots with a 100~m radius, and there are five hotspots with fixed locations, as illustrated {in} Figure~\ref{fig:hotspotScenario}. This positioning aims at modeling different situations: close and far from a \ac{eNB} and between multiple \acp{eNB}.

\begin{algorithm}[h!]
    \SetKwInOut{Input}{Input}
    \SetKwInOut{Output}{Output}

		\textbf{Initialize}
		
		\For{$s \in S, a \in A$}
		{
		Initialize Q-table with all ~$Q(s,a)$ equal to 0.
		}
	
		Estimate initial state~$s$.
		
		\textbf{Learning:}
		
		\textbf{Loop}
		
		{
			Generate random number $r \in U(0,1)$
		}
		
    \eIf{$r < \epsilon$}
      {
        Randomly choose action ~$a \in A$ \;
      }
      {
        Choose action $a \in A$ according {to Equation \ref{eq:estrategia-qlearning};} 
      }
			
		Execute action $a$\; 
		
		Receive immediate reward~$r$, SINR$_{{\rm m\acute{e}dia}}$\; 
		
		Observe next state~$s{'} \in S$\; 
		
		Update Q-table according to {Equation \ref{eq:q_values_pseudocode};}
		
		$s = s{'}$
		
		\textbf{end loop}
			
    \caption{Pseudo-code of the proposed \textit{Q-learning} algorithm.}\label{alg:pseudo_ql}
\end{algorithm}

As described {in} Section \ref{sec:interferenceHotspot}, this paper uses the term hotspot to describe regions with high user density. It does not mean a new \ac{AP}. Hence, hotspot users do not have a dedicated \ac{AP}. All users are served by the closest \ac{eNB}.

Table \ref{tab:scenarios-ql} presents the difference between the scenarios used to evaluate the \ac{QL} algorithm. \textbf{Scenario 1} has 60 users distributed uniformly and 10 users on each hotspot. A total of 52 \acp{RB} are allocated for the common sub-band, and 16 \acp{RB} are allocated to the private sub-band. \textbf{Scenario 2} has the double of users on each hotspot and 2/3 of the uniform-distributed users
when compared to Scenario 1. Hence, there are 40 users distributed uniformly and 20 users on each hotspot, which results in a higher number of users in the system. A total of 28 \acp{RB} are allocated for the common sub-band, and 24 \acp{RB} are allocated to the private sub-band. Therefore, the available bandwidth for users allocated in the common sub-band is significantly smaller. Other simulation parameters are summarized {in} Table~\ref{tab:paramQL}.

\begin{specialtable}[H]
\caption{Evaluations scenarios for proposed solution.}
\label{tab:scenarios-ql}

\begin{tabular}{m{13.3cm}}  

\toprule
\textbf{Scenario 1}                 \\
\midrule                                             
60 users uniformly distributed      \\
\midrule
10 users on each hotspot            \\
\midrule
BandwidthDistribution: 52/16        \\
\midrule
\textbf{Scenario 2}                  \\
\midrule                                             
40 users uniformly distributed      \\
\midrule
20 users on each hotspot            \\
\midrule
BandwidthDistribution: 28/24        \\
\bottomrule

\end{tabular}
\end{specialtable}
\vspace{-6pt}

\begin{specialtable}[H]
\caption{Simulation parameters for the \ac{QL} algorithm evaluation.}
\label{tab:paramQL}

\begin{tabular}{p{6.5cm}p{6.5cm}}

\toprule
\textbf{Parameter}              & \textbf{Value}            \\
\midrule
Bandwidth (\acsp{RB})           & 100                       \\     
\midrule                                             
UE distribution                 & Uniform / Hotspots        \\
\midrule
Number of \acsp{UE}             & 110 / 140                 \\
\midrule
Distance between \acsp{eNB} (m) & 1000                      \\
\midrule
Scheduling algorithm            & \textit{Proportional Fair}\\
\midrule
Simulation duration             & 60000 subframes            \\
\midrule
Channel model                   & Friis model                \\
\midrule
Error model                     & MIESM                     \\
\midrule
\acs{UE} mobility               & No mobility               \\
\midrule
Traffic model                   & Non-GBR TCP-based         \\
                                & Video (Buffered Stream)   \\
\bottomrule

\end{tabular}
\end{specialtable}

{A hotspot (HS) begins to transmit and receive data every 10 s to evaluate the algorithm's ability to adapt.} From 0 to 10 s, only the uniformly distributed users are active. Between 10 and 20 s, one of the hotspots is also transmitting/receiving data. From 30 s, another hotspot becomes active, and so forth. The change on the scenario during simulation is presented {in} Table~\ref{tab:hotspotsAppearance}. Before discussing the results, the {following} section details how data are collected on {the} ns-3 simulator to enable the \ac{QL} calculations.

\begin{specialtable}[H]
\setlength{\tabcolsep}{11.9mm}
\caption{Active users at each simulation interval.}
\label{tab:hotspotsAppearance}

\begin{tabular}{ccc}  

\toprule
\textbf{Interval}  &  \textbf{Scenario 1 (UEs)}  &  \textbf{Scenario 2 (UEs)}        \\
\midrule
0 to 10 s  &  60 + 0 on HS  &  40 + 0 on HS   \\     
\midrule                                             
10 to 20 s &  60 + 10 on HS &  40 + 20 on HS  \\
\midrule
20 to 30 s  &  60 + 20 on HS &  40 + 40 on HS \\
\midrule
30 to 40 s  &  60 + 30 on HS &  40 + 60 on HS \\
\midrule
40 to 50 s  &  60 + 40 on HS &  40 + 80 on HS \\
\midrule
50 to 60 s  &  60 + 50 on HS &  40 + 100 on HS\\
\bottomrule

\end{tabular}
\end{specialtable}

\subsection{Data Collection on ns-3}

Researchers using ns-3 can use a mechanism called \textit{tracing} to capture any data generated by simulations. With this subsystem, the user can connect a function to any \textit{trace source}, which outputs the data to the connected function when a condition is satisfied (e.g., every time a user transmits). This paper uses the trace source \textit{ReportCurrentCellRsrpSinr}. It returns linear \ac{SINR} values for all users from a specific \ac{eNB} at each \ac{LTE} subframe (every 1~ms). This trace source is implemented in the PHY layer from the \ac{LTE} module. For the throughput, the trace source \textit{RxPDU} notifies the amount of received \acp{PDU} by the \ac{RLC} entity.

The function connected to the first trace source stores \ac{SINR} values from all \acp{eNB} and, every 40~ms, calculates the average \ac{SINR} of the system on that interval (accumulated sum divided by {the} number of samples). Similarly, the second trace source is connected to a function that stores the data in text files. These files contain the number of \acp{PDU} and bytes transmitted and received. The throughput is calculated by dividing the total amount of received bits by the transmission interval in seconds. 

\subsection{Proof-of-Concept Simulation Results}
\label{sec:analise_resultados}

The results are presented using \ac{SINR} and throughput. Moreover, users are split in three groups: all users, users in hotspots, and users outside hotspots. \mbox{Figures \ref{fig:sinrGain1} and \ref{fig:sinrGain2}} present the \ac{SINR} percentage gains for all groups of users. Figures \ref{fig:avgtputQL1} and \ref{fig:avgtputQL2} present throughput results in Mbps and Figures \ref{fig:tputgain1} and \ref{fig:tputgain2} present the throughput percentage gain.

Given the wide range that \ac{SINR} is reported in linear scale, it is more practical to calculate the gain {on} a logarithmic scale. Hence, it is executed using dB values, as {follows}:

\begin{equation}
Gain = \left( \frac{SINR_{QL}}{SINR_{noQL}}  - 1 \right) \times 100\%
\end{equation}

\noindent where $SINR_{QL}$ is the system average $SINR$ in dB when our solution is applied, and $SINR_{noQL}$ is the system average $SINR$ without it. Thus, the $Gain$ measures how our proposal improves the SINR in {the} dB scale.

According to Figure \ref{fig:sinrGain1}, all intervals and groups of users present relevant gain, even when there are no active hotspots (0 to 10 s). Additionally, as the number of active hotspots increase{s}, the \ac{SINR} degrades more severely, and the gain obtained tends to grow, since \ac{QL} operates to improve the \ac{SINR}. Hence, the two intervals with more active hotspots present over 80\% gain. In addition, for most intervals, the gain is greater for hotspot users, which is interesting, {considering} the algorithm was not configured to act based on the performance of these users.

% start a new page without indent 4.6cm
\clearpage
\end{paracol}
\nointerlineskip
\begin{figure}[H]
\widefigure
\includegraphics[width=\textwidth]{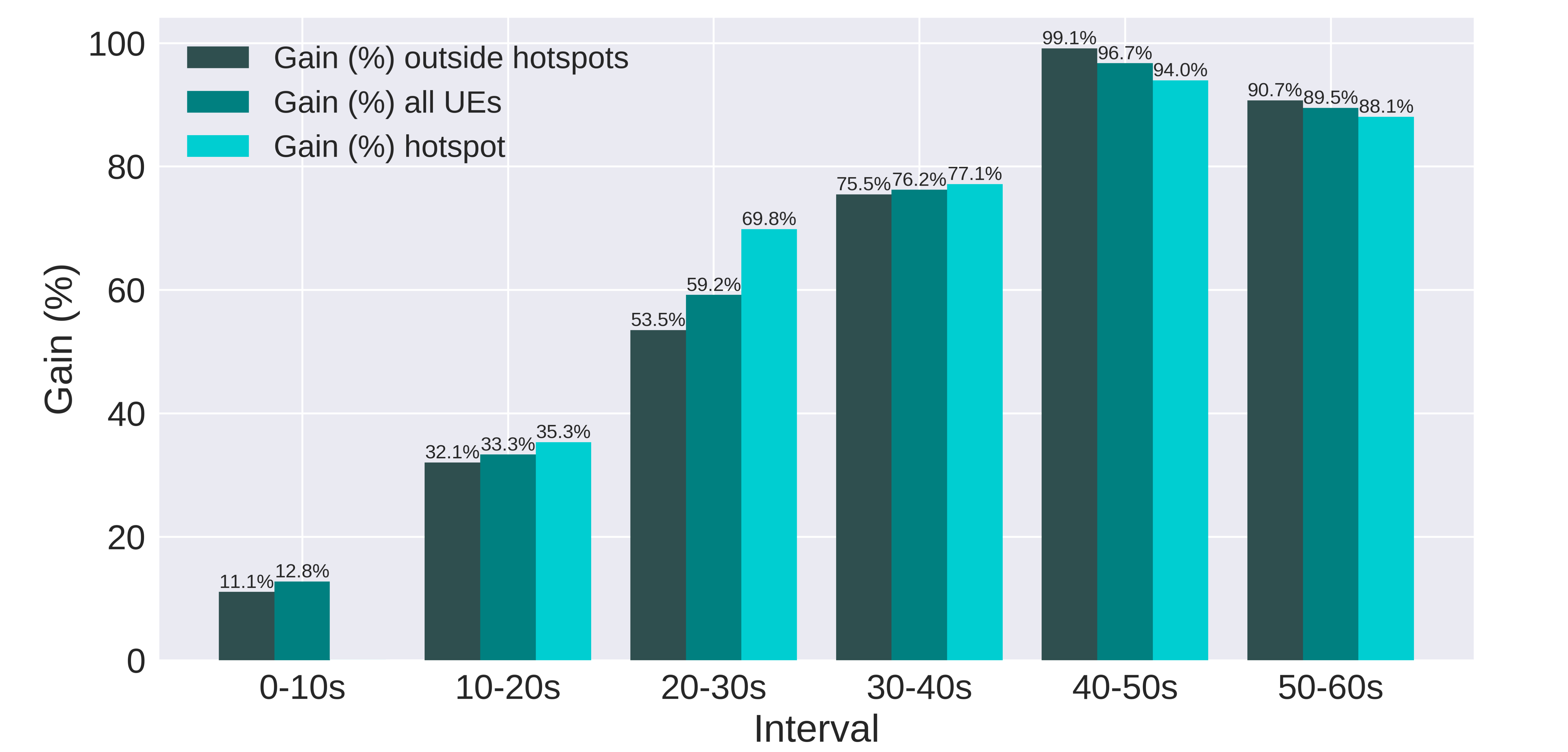}
\caption{\label{fig:sinrGain1} {Scenario 1 of the proposed solution: SINR Gain (\%) for each group of users.}}
%\vspace{0.4cm}

\end{figure}
\begin{paracol}{2}
%\linenumbers
\switchcolumn

Figure \ref{fig:avgtputQL1} presents throughput results for the same scenario. When the \ac{QL} is not active, RsrqThreshold is fixed on 32. In this plot, three important results are combined: {the} throughput of users outside the hotspots (bottom bars), {the} throughput of hotspot users (top bars), and the average throughput of all users (combination of both bars). Each group of two parallel bars corresponds to a 10 s interval during the simulation (Table \ref{tab:hotspotsAppearance}), and the bars on the left present the results when the \ac{QL} is not active. The plot shows that the \ac{SINR} gain resulted {in} throughput gain for all groups of users. However, the gain for users outside the hotspots is smaller when compared to hotspot users.

% start a new page without indent 4.6cm
%\clearpage
\end{paracol}
\nointerlineskip
\begin{figure}[H]
\widefigure
\includegraphics[width=\textwidth]{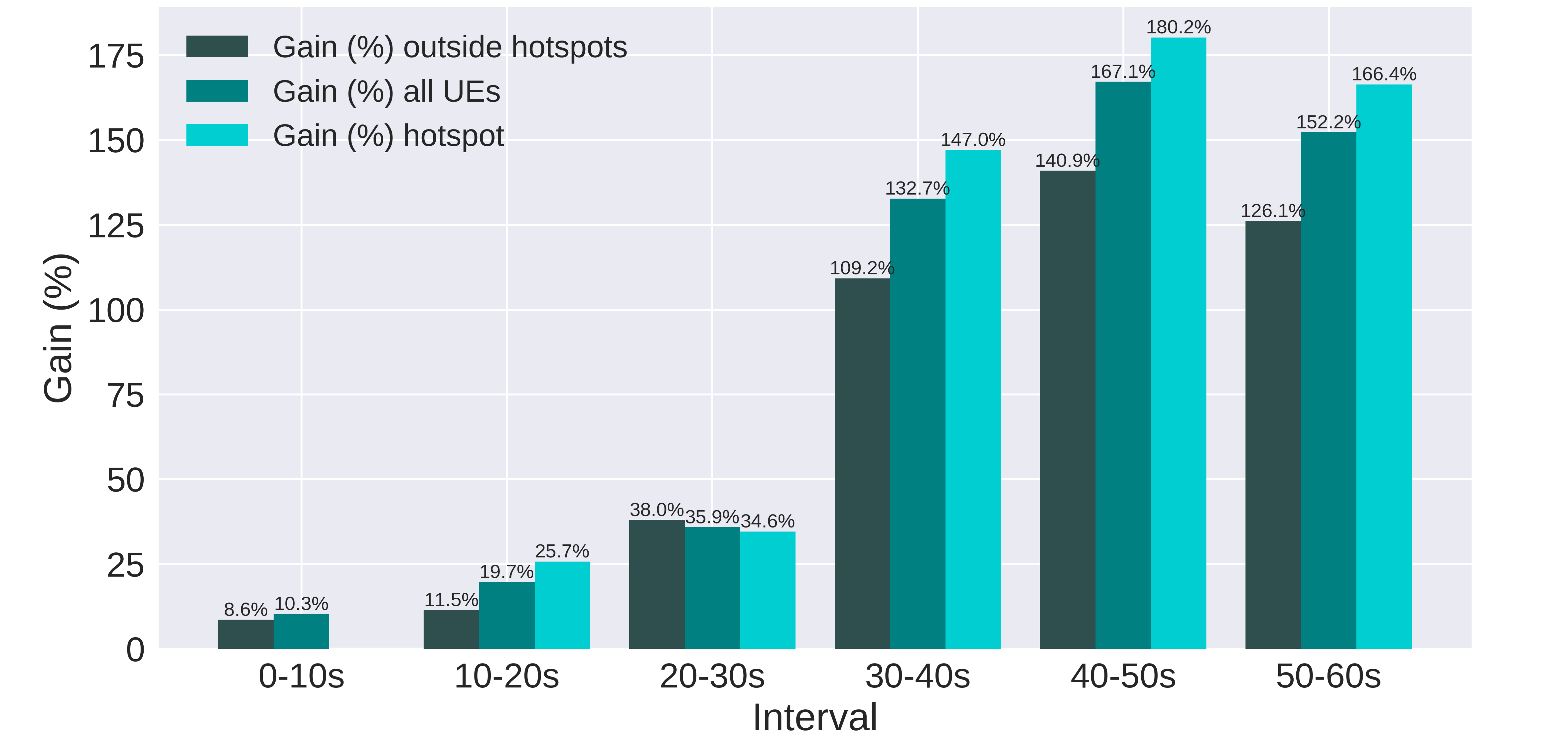}
\caption{\label{fig:sinrGain2} {Scenario 2 of the proposed solution: SINR Gain (\%) for each group of users.}}
%\vspace{0.3cm}

\end{figure}
\begin{paracol}{2}
%\linenumbers
\switchcolumn

Furthermore, Scenario 2 investigates the algorithm's performance with more users allocated in the hotspots and with different configuration{s} for the \textit{BandwidthDistribution} parameter, which also has {a} significant impact on the evaluated scenario (Section \ref{sec:2kfactorial}). This scenario allocates less bandwidth for the common sub-band and more bandwidth for the private sub-band.

Figure \ref{fig:sinrGain2} presents the \ac{SINR} gain for Scenario 2. Similar to Scenario 1, all groups of users have significant gain, and it is approximately doubled for the last three intervals, compared to the previous scenario. This is mainly due to the double amount of users on each hotspot of Scenario 2, which degrades the \ac{SINR} more severely. Since the \ac{SINR} is the target metric of the proposed algorithm, the gain is also doubled, showing that the \ac{QL} is efficient in improving the \ac{SINR}, even if the sub-bands are configured differently.

\begin{figure}[H]
\includegraphics[width=0.85\linewidth]{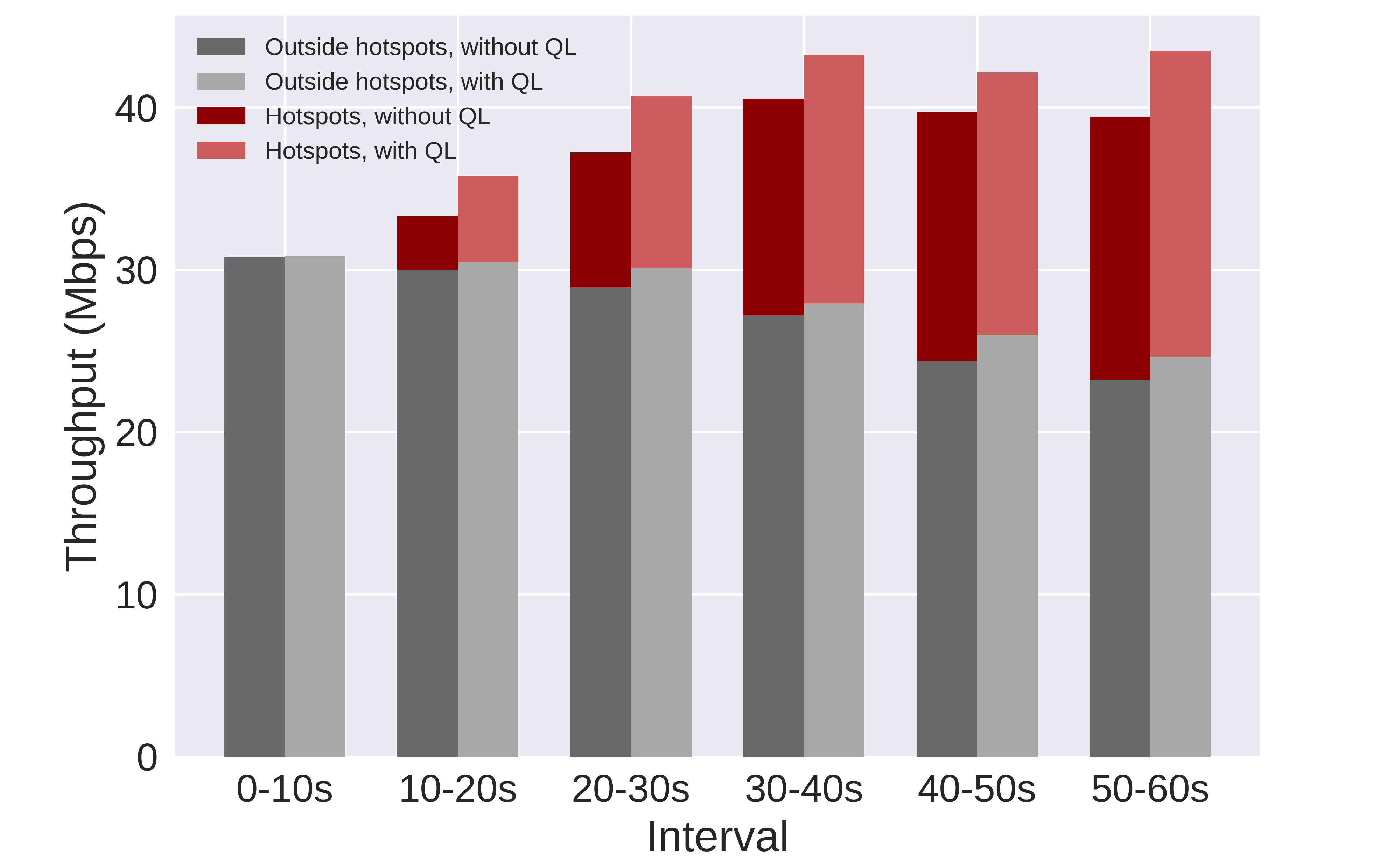}
\caption{\label{fig:avgtputQL1} {Scenario 1 of the proposed solution: throughput results for each group of users, with and without Q-Learning.}}
%\vspace{0.3cm}

\end{figure}

\begin{figure}[H]
\includegraphics[width=0.85\linewidth]{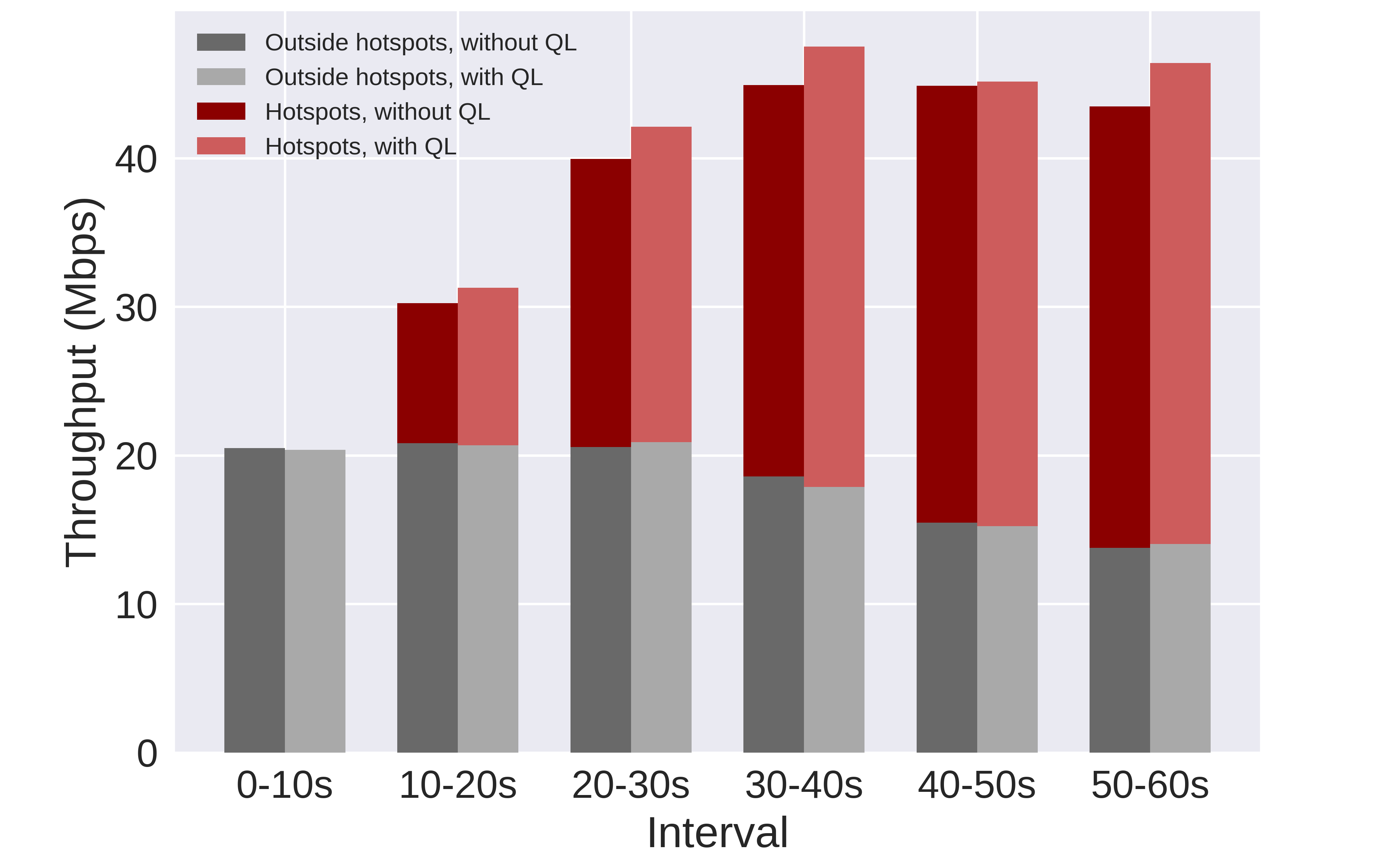}
\caption{\label{fig:avgtputQL2} {Scenario 2 of the proposed solution: throughput results for each group of users, with and without Q-Learning.}}
%\vspace{0.3cm}

\end{figure}

Figure \ref{fig:avgtputQL2} presents throughput results for Scenario 2. There is {a} relevant gain for the average throughput and for hotspot users. However, the gain is minimal to users outside of hotspots. On the third interval (30 to 40 s), these users suffer a small loss, but the gain from hotspot users compensate{s}, resulting in {a} gain in the system average throughput.

Summarizing, the throughput percentage gain is presented in Figures \ref{fig:tputgain1} and \ref{fig:tputgain2}, for scenarios 1 and 2, respectively. The smallest gain obtained for hotspot users was 5.4\% in scenario~1 and 1.7\% in scenario~2. On the other hand, the highest gain for these users was 59.8\% in scenario~1 and 12.5\% in scenario~2. Users outside the hotspots presented loss in some intervals for scenario~2. Nevertheless, all intervals presented gain for hotspot users. These users were not present in the plots for the first interval because there were no active hotspots users. The worst and best percentage gain for each scenario and group of users are summarized in Table \ref{tab:summay_results_QL}.

% start a new page without indent 4.6cm
%\clearpage
\end{paracol}
\nointerlineskip

\begin{specialtable}[H]
\setlength{\tabcolsep}{5.2mm}
\widetable 
\caption{{Best and worst results for SINR and throughput of the Q-Learning solution, for each group of users.}}
\label{tab:summay_results_QL}

\begin{tabular}{lllll}
\toprule
& \textbf{Best SINR Gain} &  \textbf{Worst SINR Gain}  & \textbf{Best Tput Gain} & \textbf{Worst Tput Gain}  \\
\midrule
\textbf{Scenario 1} & & & & \\
\midrule
Hotspot users & 94\%  & 35.3\%   & 59.8\%    & 5.4\%   \\
\midrule
Outside hotspots & 99.1\% & 11.1\% & 6.5\% & 0.2\% \\
\midrule
All Users & 96,7\% & 12.8\% & 10.3\% & 0.2\% \\
\midrule
\textbf{Scenario 2} & & & & \\
\midrule
Hotspot users & 180.2\%   & 25.7\%   & 12.5\%    & 1.7\% \\
\midrule
Outside hotspots & 140.9\% & 8.6\% & 1.9\% & $-$3.8\%\\
\midrule
All Users & 167.1\% & 10.3\% & 6.7\% & $-$0.6\% \\
\bottomrule
\end{tabular}
\end{specialtable}
\begin{paracol}{2}
%\linenumbers
\switchcolumn

\begin{figure}[H]
\includegraphics[width=\linewidth]{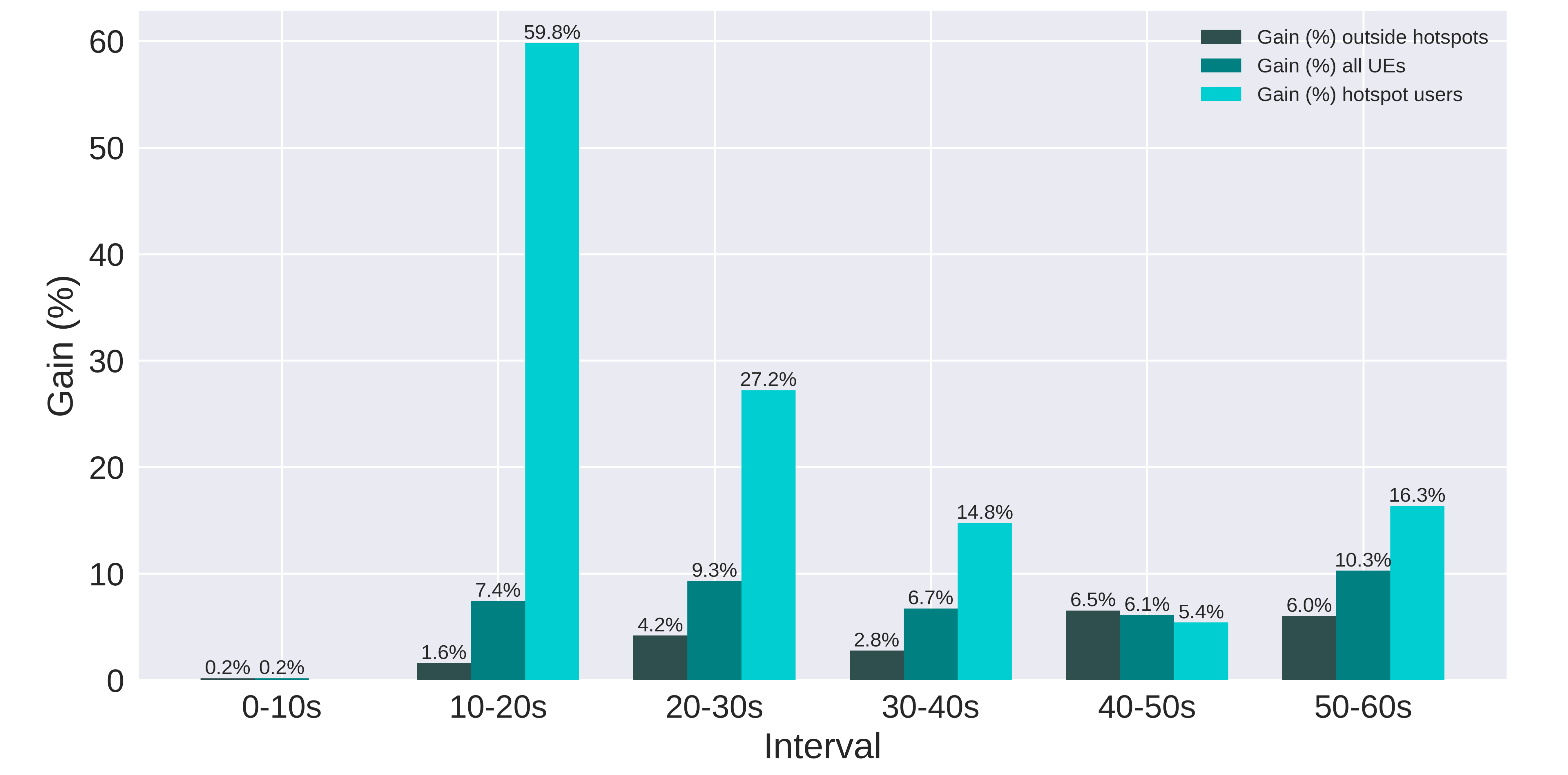}
\caption{\label{fig:tputgain1} {Scenario 1 of the proposed solution: throughput gain (\%) for each group of users.}}
\vspace{0.3cm}

\end{figure}

\begin{figure}[H]
\includegraphics[width=\linewidth]{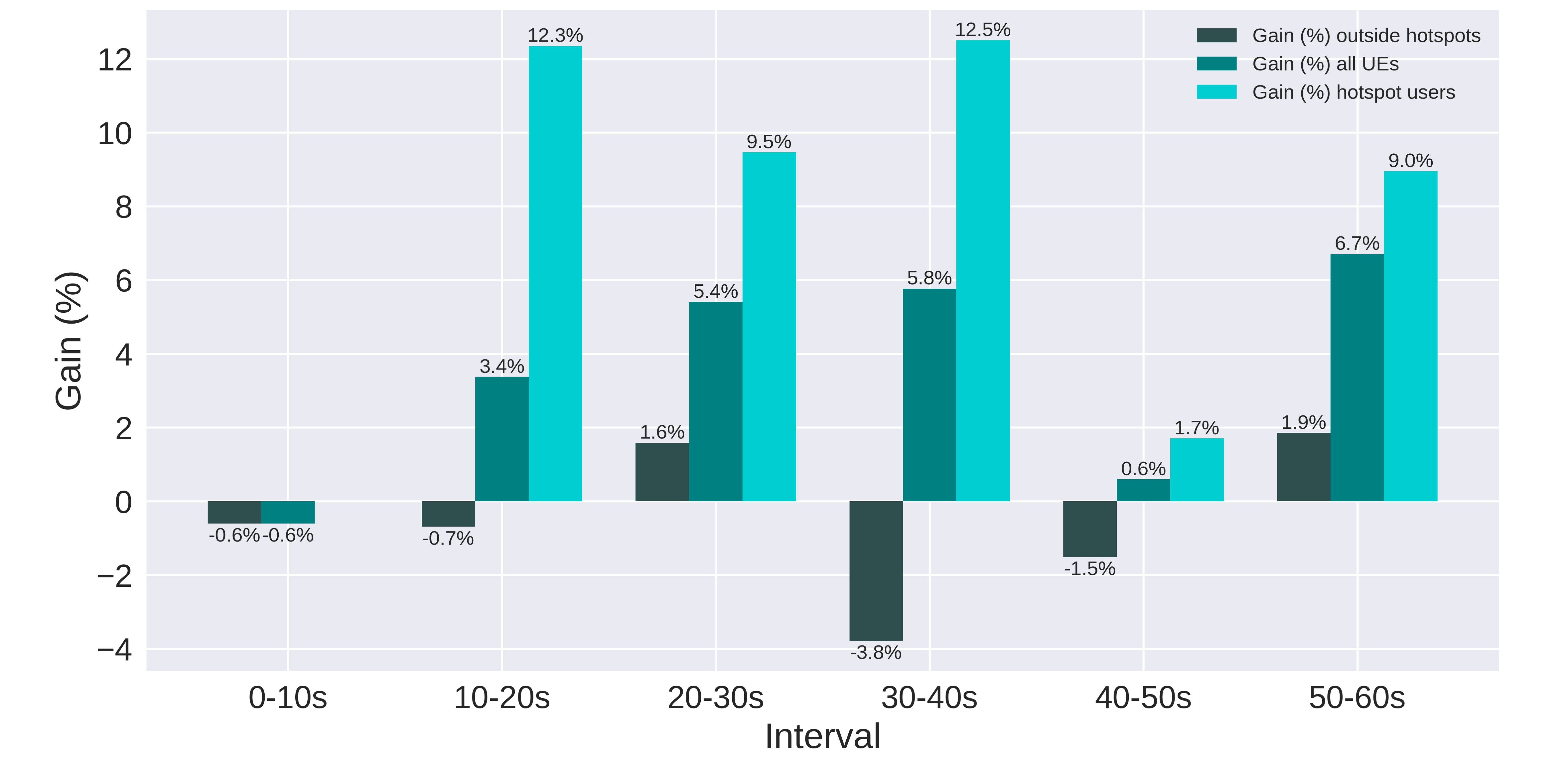}
\caption{\label{fig:tputgain2} {Scenario 2 of the proposed solution: throughput gain (\%) for each group of users.}}
\vspace{0.3cm}

\end{figure}

\section{Conclusions}

In recent years, mobile networks have experienced an increase {in} the {number} of scenarios with challenging requirements, {especially} due to rapid growth in traffic demand and {the} arrival of new services, such as \ac{eMBB}. Likewise, user behavior has changed, creating scenarios that are hard to predict. This paper discussed the challenges related to a dynamic hotspot scenario and how to mitigate the resulting performance loss. A solution was presented to coordinate \ac{ICI} using \ac{FFR}-based techniques. The algorithm dynamically regulates the \ac{Strict FR} parameters, using \acf{QL}.

The results indicate that the solution effectively optimizes the \ac{SINR} in the evaluated scenarios, which also resulted {in} throughput gain, even though this {relationship} is not always true. Depending on the scenario, the \ac{SINR} gain may only result in a decrease of the \ac{PLR}, which is also important for the overall performance of the system.

In some cases, the gain was higher for hotspot users, even though the algorithm was not configured to act based on their performance. Besides, even in these cases, the algorithm also improved the overall system performance. The results also indicate a tendency of greater \ac{SINR} gain when the scenario suffers with more interference. The higher \ac{SINR} gain obtained was 180\% for hotspot users, 167.1\% for average \ac{SINR}, and 140\% for users \mbox{outside hotspots.}

\section{Future Works}

The development of research usually involves various choices in order to define the scope of the paper. As a consequence, some unexplored possibilities and possible improvements are listed below as an encouragement to future endeavors. 

\begin{itemize}\itemsep0em
    
    \item Use other metrics to define the \ac{QL} states and actions, such as the BandwidthDistribution. As presented {in} Section \ref{sec:q-learning}, these metrics can also be the combination of different variables. As a result, the algorithm is expected to improve in flexibility and efficiency when using two parameters, adapting to a {broader} set of scenarios. For example, the throughput could be included as part of the reward;
    
    \item Expand proposed scenarios: vary {number} of users, add mobility, vary{ing} number and location of hotspots (which could also appear in random locations), and test different values for BandwidthDistribution; 
    
    \item Evaluate the system using different metrics, such as the relation between convergence speed, the amount of state/action pairs, or the \acf{PLR};
    
    \item Apply the solution on a system without isotropic antennas, such that transmission is made in sectors within a cell;
    
    \item Apply the presented solution for 5G \acf{NR} or the \ac{LTE} \ac{UL}, given that the interference in the \ac{UL} has different characteristics, when compared to \ac{DL};
    
    \item Provide a similar solution, replacing the RL algorithm for, e.g., multi-armed bandit, providing a simpler solution.
    
\end{itemize}

%%%%%%%%%%%%%%%%%%%%%%%%%%%%%%%%%%%%%%%%%%
\vspace{6pt} 

%%%%%%%%%%%%%%%%%%%%%%%%%%%%%%%%%%%%%%%%%%
%% optional
%\supplementary{The following are available online at \linksupplementary{s1}, Figure S1: title, Table S1: title, Video S1: title.}


%\sampleavailability{}

%%%%%%%%%%%%%%%%%%%%%%%%%%%%%%%%%%%%%%%%%%
\authorcontributions{Conceptualization, I.D.d.R., V.A.d.S.J.; data curation, I.D.d.R.; formal analysis, I.D.d.R.; supervision, V.A.d.S.J.; visualization, I.D.d.R.; writing---original draft, I.D.d.R.; writing---reviewing and editing, V.A.d.S.J. All authors have read and agreed to the published version of the manuscript.}

%%%%%%%%%%%%%%%%%%%%%%%%%%%%%%%%%%%%%%%%%%
\funding{This study was financed in part by the Coordena\c{c}\~{a}o de Aperfei\c{c}oamento de Pessoal de N\'{i}vel Superior---Brasil (CAPES)---Finance Code 001.}

\institutionalreview{Not applicable.}%In this section, please add the Institutional Review Board Statement and approval number for studies involving humans or animals. Please note that the Editorial Office might ask you for further information. Please add ``The study was conducted according to the guidelines of the Declaration of Helsinki, and approved by the Institutional Review Board (or Ethics Committee) of NAME OF INSTITUTE (protocol code XXX and date of approval).'' OR ``Ethical review and approval were waived for this study, due to REASON (please provide a detailed justification).'' OR ``Not applicable'' for studies not involving humans or animals. You might also choose to exclude this statement if the study did not involve humans or animals.}

\informedconsent{Not applicable.}%Any research article describing a study involving humans should contain this statement. Please add ``Informed consent was obtained from all subjects involved in the study.'' OR ``Patient consent was waived due to REASON (please provide a detailed justification).'' OR ``Not applicable'' for studies not involving humans. You might also choose to exclude this statement if the study did not involve humans.

%Written informed consent for publication must be obtained from participating patients who can be identified (including by the patients themselves). Please state ``Written informed consent has been obtained from the patient(s) to publish this paper'' if applicable.}

\dataavailability{Not applicable. }%In this section, please provide details regarding where data supporting reported results can be found, including links to publicly archived datasets analyzed or generated during the study. Please refer to suggested Data Availability Statements in section ``MDPI Research Data Policies'' at \url{https://www.mdpi.com/ethics}. You might choose to exclude this statement if the study did not report any data.} 

%%%%%%%%%%%%%%%%%%%%%%%%%%%%%%%%%%%%%%%%%%
\acknowledgments{The proof of concept simulations provided by this paper was supported by the High Performance Computing Center at UFRN (NPAD/UFRN).}

%%%%%%%%%%%%%%%%%%%%%%%%%%%%%%%%%%%%%%%%%%
\conflictsofinterest{The authors declare that they have no competing interests.}

\end{paracol}
\reftitle{References}

%%%%%%%%%%%%%%%%%%%%%%%%%%%%%%%%%%%%%%%%%%
%% optional

%% for journal Sci
%\reviewreports{\\
%Reviewer 1 comments and authors’ response\\
%Reviewer 2 comments and authors’ response\\
%Reviewer 3 comments and authors’ response
%}

%%%%%%%%%%%%%%%%%%%%%%%%%%%%%%%%%%%%%%%%%%
\end{document}